\documentclass[aps,prx,twocolumn, longbibliography, superscriptaddress]{revtex4-1}

\usepackage{color}
\usepackage{array}
\usepackage{amsmath}
\usepackage{amssymb}
\usepackage{wasysym}
\usepackage{bm}
\usepackage{graphicx}
\usepackage{txfonts}
\usepackage{epsfig}
\usepackage{hyperref}
\usepackage{multirow}
\usepackage{verbatim}
\usepackage{esint,hyperref}
\usepackage{makecell}

\newcommand{\bra}[1]{\langle \, #1 \,|}
\newcommand{\ket}[1]{|\, #1 \, \rangle}
\newcommand{\bket}[2]{\langle \, #1 \,|\, #2 \, \rangle}
\newcommand{\boket}[3]{\langle\, #1 \,|\, #2 \,|\, #3 \,\rangle}
\newcommand{\ketbra}[1]{\ket{#1} \bra{#1}}

\newcommand{\be}{\begin{equation}}
\newcommand{\ee}{\end{equation}}
\newcommand{\bc}{\begin{center}}
\newcommand{\ec}{\end{center}}
\newcommand{\nin}{\noindent}

\newcommand{\non}{\nonumber}

\begin{document}
\title{Quantum Origami: Transversal Gates for Quantum Computation \\ and Measurement of Topological Order }
\author{Guanyu Zhu}
\affiliation{Joint Quantum Institute, NIST/University of Maryland, College Park, MD 20742 USA}
\author{Mohammad Hafezi}
\affiliation{Joint Quantum Institute, NIST/University of Maryland, College Park, MD 20742 USA}
\affiliation{Department of Physics and Astronomy, University of Maryland, College Park, MD 20742 USA}
\affiliation{Department of Electrical and Computer Engineering and Institute for Research in Electronics and Applied Physics,
University of Maryland, College Park, MD 20742, USA}
\author{Maissam Barkeshli}
\affiliation{Joint Quantum Institute, NIST/University of Maryland, College Park, MD 20742 USA}
\affiliation{Department of Physics and Astronomy, University of Maryland, College Park, MD 20742 USA}
\affiliation{Condensed Matter Theory Center, University of Maryland,
  College Park, Maryland 20742, USA}

\begin{abstract}
In topology, a torus remains invariant under certain non-trivial transformations known as modular transformations.  In the context of topologically ordered quantum states of matter supported on a torus geometry in real space, these transformations encode the braiding statistics and fusion rules of emergent anyonic excitations and thus serve as a diagnostic of topological order. Moreover, modular transformations of higher genus surfaces, e.g.~a torus with multiple handles, can enhance the computational power of a topological state, in many cases providing a universal fault-tolerant set of gates for quantum computation. However, due to the intrusive nature of modular transformations, which abstractly involve global operations, physical implementations of them in local systems have remained elusive. Here, we show that by engineering an effectively folded manifold corresponding to a multi-layer topological system, modular transformations can be applied in a single shot by independent local unitaries, providing a novel class of transversal logical gates for fault-tolerant quantum computation. Specifically, we demonstrate that multi-layer topological states with appropriate boundary conditions  and twist defects allow modular transformations to be 
effectively implemented by a finite sequence of local SWAP gates between  the layers. We further provide methods to directly measure the modular matrices, and thus the fractional statistics of anyonic excitations, providing a novel way to directly measure topological order. A more general theory of transversal gates and the deep connection to anyon symmetry transformation and symmetry-enriched topological orders is also discussed. 
\end{abstract}

\maketitle

\section{Introduction}

The last few decades in condensed matter physics, starting with the discovery of the quantum Hall effect, 
have yielded remarkable progress in the understanding of the possible quantum states of matter. In particular,
at zero temperature, many-body quantum systems can form distinct phases of matter that are distinguished by
their topological order \cite{Wen:1990tm}. Topologically ordered states, such as fractional quantum Hall (FQH) states and quantum spin 
liquids, exhibit a host of remarkable properties, including quasiparticles with exotic, possibly non-Abelian, 
exchange statistics, fractional charges, robust topology-dependent ground state degeneracies, and protected 
gapless edge modes \cite{Nayak:2008dp, wen04}. The intricate patterns of quasiparticle fusion rules and fractional statistics distinguish 
different topological orders from each other. 

The ground state degeneracy of a topologically ordered state of matter depends on the topology of the (real) space on which it is
defined \cite{Wen:1990tm}. These degeneracies are topologically protected, in the sense that no local operators can distinguish the different states, up to exponentially small corrections in system size. For this reason, topological states of matter provide one of the most promising routes
to realizing quantum error correction (QEC) and thus robust, fault-tolerant quantum computation \cite{Nayak:2008dp}. 

Topologically ordered states are useful for quantum error correction in two separate ways as summarized in Table \ref{summary_table}. One route is via a `passive' approach, where one considers a physical system described by a many-body Hamiltonian whose 
ground state is a topologically ordered state of matter \cite{Nayak:2008dp}. The topological protection then derives from cooling 
the system to temperatures much smaller than the energy gap. A second route is via an `active' approach to 
quantum error correction, where one starts with a large collection of physical qubits and continuously performs
certain `stabilizer' measurements associated with a topological code \cite{fowler2012}. These measurements project the many-body quantum
state of the qubits onto a topologically ordered state. 

While topologically protected ground state degeneracies provide a path towards realizing fault-tolerant quantum memories, 
our understanding of how to perform robust fault-tolerant quantum gates for quantum computation is limited. 
Moreover, despite the great theoretical and experimental advances of the last several decades, methods to experimentally measure fractional statistics
of quasiparticles and therefore to directly diagnose topological order in any experimental setting are extremely limited. Most experimental measurements 
that are routinely performed, such as measuring quantized electrical or thermal Hall conductivity, provide extremely limited windows into the nature of topological order. 
The purpose of this paper is to address these two issues, of experimentally measuring topological order and implementing fault-tolerant logical quantum gates. 

Topologically protected ground states can be manipulated by a set of topologically protected unitary transformations 
that act on the ground state subspace. A well-known example is when the degeneracy arises from non-Abelian anyons \cite{Nayak:2008dp}; in this case, the corresponding robust unitary transformations that act on the topologically protected subspace are generated by adiabatically braiding 
the anyons. In general, a given spatial manifold $\Sigma$ admits a group of operations known as the mapping class
group (MCG) of the manifold. Elements of the MCG are referred to as modular transformations, and consist of smooth deformations of the geometry that bring 
the manifold back to itself, modulo those that are continuously deformable to the identity map \cite{Farb_Book}. 
The braid group for $n$ anyons can be viewed as a special case of the MCG of a disk with $n$ punctures. The ground state subspace $\mathcal{H}_\Sigma$,
which arises when the system is defined on a spatial manifold $\Sigma$, forms a unitary representation of the MCG.  Therefore, the modular transformations 
acting on the manifold $\Sigma$ correspond to topologically protected unitary transformations acting on the ground-state subspace $\mathcal{H}_\Sigma$ \cite{barkeshli2016mcg}.

Remarkably, the representation of modular transformations of a torus fully characterizes the fractional statistics and fusion rules of the quasiparticle excitations of 
a topological state; therefore it can serve as an order parameter for diagnosing topological 
order \cite{VERLINDE:1988fs, moore1989, Witten1989_CMP, Wen:1990tm, Nayak:2008dp,wang2008,zhang2012,wen2012,barkeshli2014SDG}. 
On the other hand,  the representation of a MCG of a higher genus surface can typically enhance the computational power that can be achieved 
by braiding the anyons alone \cite{wang2008}. For example, the representation of the braid group for anyons in an Abelian topological state is 
always one-dimensional and thus the braiding transformations only lead to an overall phase in the wave function. On the other hand, the representation of the MCG 
on a higher genus is always multi-dimensional and leads to nontrivial unitary transformations corresponding to fault-tolerant logical gates. 
In some cases of non-abelian topological states, such as for the Ising anyon topological state, the braid group of non-abelian anyons does not provide
a universal gate set for fault-tolerant quantum computation, while access to the MCG on a high genus surface can provide a 
universal fault-tolerant logical gate set \cite{bravyi200universal, freedman2006Ising, barkeshli2016mcg}. 

We see that the ability to experimentally measure the modular matrices in a given state of matter would imply the ability to directly measure
fractional statistics and thus fully diagnose topological order. The ability to implement modular transformations can further enhance the ability
to perform fault-tolerant quantum computation. It is thus of fundamental importance to develop methods to implement and measure modular transformations. 

Previous schemes for implementing modular transformations in physical systems involve intrusive techniques such as adiabatically 
varying the  geometry or interactions \cite{You:2015vn, barkeshli2016mcg, Barkeshli:2016dn,breuckmann2017}, braiding twist 
defects \cite{Barkeshli:2013da}, topological charge measurements \cite{barkeshli2016mcg, Bonderson:2010vd}, or via global
rotations of a torus \cite{zhang2012,wen2012,luo2017}. Crucially, all of these methods require a time overhead that scales polynomially 
with system size; that is, the time scale to carry out these operations is infinite in the thermodynamic limit. Moreover, the methods 
presented in Refs.~\cite{You:2015vn, zhang2012,wen2012,luo2017,breuckmann2017} cannot be applied in a planar geometry 
with local interactions, even in principle; as such, they are not feasible for realistic experimental settings. 

Motivated by recent experimental progress in implementing error correcting codes in superconducting 
qubit systems \cite{Magesan:2015bg, Barends:2014fub, egrave:2015dx, Gambetta:2017iu, Bonesteel:2012fl, Wosnitzka:2016jt, Lekitsch:2015ua} 
and also the unprecedented level of local control in quantum gas microscopes \cite{Islam:2015cm}, it is timely to consider 
whether the seemingly abstract mathematical operations of modular transformations can be effectively implemented in real 
physical systems with local interactions.

\begin{table}
\begin{tabular}{c||c|c}

  & Topological matter  &  Topological QECC   \\     

\hline
\multirow{2}{*}{Hilbert space $\mathcal{H}_\Sigma$}  &deg. ground state  & \multirow{2}{*}{code space }  \\
& subspace  &        \\                     
\hline
Protection scheme                         & passive                                      & active           \\ 
 \hline                                                                                                                                                    
 Representation of               & \multicolumn{2}{c}{ logical gates   }                    \\
 \cline{2-3}
 modular transformations  & \makecell{ diagnose \\ topological order}   & \makecell{ verify \\topological order} \\
\hline          
\end{tabular}
\caption{Dictionary about the connection between several concepts: topological phases of matter, topological quantum error correcting codes (QECC), manifold topology, and fault-tolerant logical gates.}
\label{summary_table}
\end{table}

In this work, by folding manifolds in various ways, \textit{quantum origami}, we demonstrate how modular transformations can be physically 
implemented in a single shot, i.e., with constant time overhead, and in a planar geometry with fully local interactions. In particular, we 
consider a multi-layer topological state, i.e., multiple copies of the same topological state aligned vertically in real space, and with 
appropriate boundary conditions and twist defects connecting different layers \cite{Barkeshli:2010bo, Barkeshli:2012kw, Barkeshli:2013da, Barkeshli:2014by}. 
Our scheme uses local pairwise SWAP gates between certain layers. Transformations of this form are \textit{transversal logical gates} (TLG), 
as they can be performed in one shot with a simultaneous unitary transformation independently on each subsystem (labeled by $j$) under a transversal partitioning of the entire system, i.e.,
$\overline{U}=\prod_j U_j$, following the strict   definition in Ref.~\cite{Eastin:2009cj}. We emphasize that in this paper, sites on all layers aligned vertically are merged into a single subsystem and labeled by $j$.  Since the boundary of these layers are glued together, we should treat all layers as a single code block, in contrast to the more commonly encountered case where transversal CNOTs are applied between two code blocks corresponding to two independent layers.  
In particular, we can see that our gates do not couple different subsystems in the same code block, which is the other equivalent definition for TLGs \cite{Eastin:2009cj}.

TLGs are inherently fault tolerant since the errors cannot be propagated to other subsystems \cite{Eastin:2009cj, Beverland:2016bi}.  Their  one-shot nature dramatically speeds up the time to perform logical operations, as they avoid the overhead of other schemes, such as those based on braiding, which grow linearly with the system size (or code distance). As such, TLGs are highly
coveted for fault-tolerant quantum computation, however only a limited class of them are known to date \cite{Bravyi:2013dx, Pastawski:2014, OConnor:2018}, and a number of no-go theorems prohibit their existence under general assumptions \cite{Eastin:2009cj, Beverland:2016bi}. Our results allow us to dramatically extend the space of known 
TLGs in topological codes -- especially non-Abelian ones -- by providing a novel geometric picture for a large class of TLGs. Remarkably, we 
demonstrate how to bypass the assumptions of previous no-go theorem in Ref.~\cite{Beverland:2016bi}, and demonstrate the necessary ingredients that allow a wide class of TLGs to be implemented in topological codes.

We emphasize that in the context of TLGs, the system does not necessarily need to physically involve multiple layers.  One can always use a single physical layer of a topological code which is equivalent to multiple virtual layers of topological phases. A well-known example is the equivalence between a single layer of color code and  two layers of toric codes \cite{Kubica:2015br}.  The virtual layer permutation symmetry considered in our construction is a special example of onsite topological symmetries in the context of symmetry-enriched topological (SET) phases \cite{barkeshli2014SDG, ChengSET2017, HeinrichSET2016}. Therefore, our construction also reveals the deep connection between symmetry-enriched topological phases and transversal logical gates.  In particular, spatial symmetries of a topological phase such as rotation and reflection symmetry can be turned into onsite symmetry by folding (\textit{quantum origami}), and the onsite symmetry plays the role of TLGs.

We note that our results are closely related to several previous studies on the relation between fault-tolerant logical gate, gapped domain walls, and anyon symmetries in the context of Abelian topological stabilizer codes \cite{Yoshida2017, Yoshida2015, Webster2018}. Our work builds on these results by utilizing more general constructions involving twist defects, and demonstrating how modular transformations can always be implemented as TLGs for Abelian and non-Abelian topological codes and phases of matter. 

We discuss an experimental platform in terms of a single-layer superconducting qubit array, which can potentially be used for realizing our proposal. The qubit array can be used via either the passive or active approaches to realizing topologically ordered states. In the passive approach, we consider a qubit array engineered with appropriate designer interactions such that the system enters a topological
phase of matter, such as a spin liquid or a FQH state, at low temperatures \cite{Roushan:2016iu, Kapit:2014vba, Kapit:2015cy, Ma:2016hh,barkeshli2015}. 
In the active approach, we consider quantum error correction with surface codes using active error syndrome measurements \cite{fowler2012}. We
further discuss various methods to measure the modular matrices in an experimental setting. These schemes can be used, for example, 
to demonstrate fractional statistics and also to experimentally diagnose and distinguish candidate topological orders,
such as $\mathbb{Z}_2$ spin liquid or double semion phases \cite{Freedman:2004ep}, in frustrated spin models.  For the near-term realization of active topological error correction codes, our measurement protocols can also be used to verify the underlying topological order in the code space. 

In order to facilitate the reading of this paper, we summarize our major achievements below and in Table I: We
(i) Provide a novel geometric picture for a large class of transversal logical gates (TLGs) in the context of both active and passive topological quantum computation. (ii) Extend the space of known TLGs in topological matter/codes – to the cases with twist defects and the cases of non-Abelian topological matter/codes, and hence circumvent the no-go theorem in Ref.~\cite{Beverland:2016bi} by introducing defects.   (iii) Propose measurement protocols and an experimental implementation in a passive topological system, to demonstrate fractional statistics and diagnose/distinguish candidate topological orders by measuring modular transformations via transversal operations. (iv) Propose an experimental implementation in an active topological code, to apply TLG and verify topological order. (v) Reveal the deep connections between topological operations (i.e., modular transformations), symmetry operations (including spatial and onsite symmetries), symmetry-enriched topological orders, and transversal logical gates in fault-tolerant quantum computation.

\begin{figure}
\includegraphics[width=1\columnwidth]{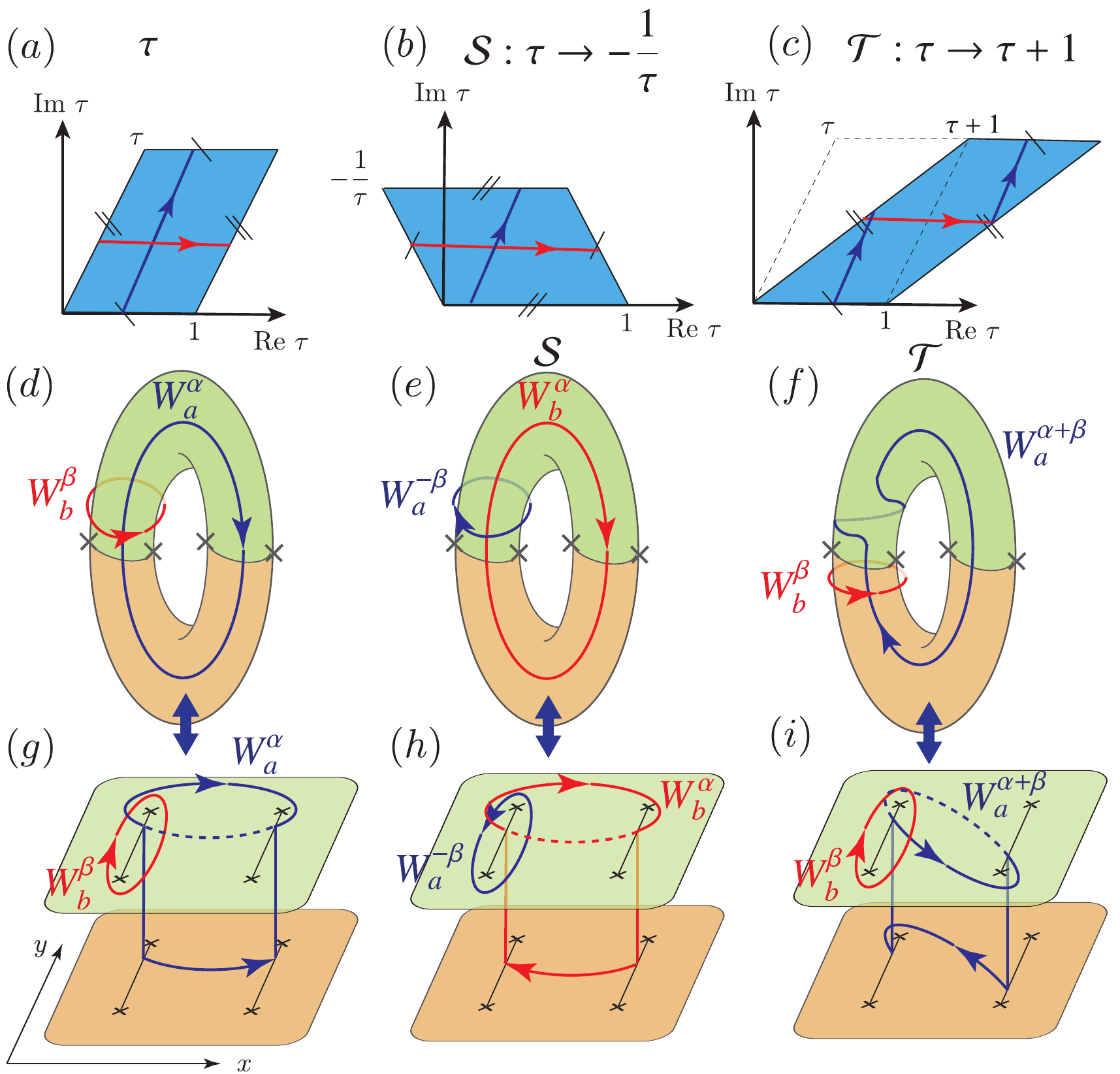}
\caption{(a-c) Definition of modular $\mathcal{S}$ and $\mathcal{T}$ transformations of a torus.  (d-i)  Non-contractible Wilson loops and their transformations under modular $\mathcal{S}$ and $\mathcal{T}$ on an effective torus realized by a double-layer topological system in the presence of genons (shown as crosses), i.e., the point defects at the end of the branch cuts (solid lines) connecting the two layers. To get a closed torus, we assume that the boundary in (g) 
has trivial topological charge. To draw $\alpha + \beta$ in (i) we use the fact that the double loop around a single genon is contractible (see Appendix \ref{append:loop-relation}).  The $\mathcal{S}$ transformation exchanges the two loops and flips the direction of $\beta$-loops  as shown in (e) and (h), while the $\mathcal{T}$ transformation performs a Dehn twist along the $\beta$-loop as shown in (f) and (i).}
\label{fig:setup}
\end{figure}

\section{Ground state degeneracies and modular transformation on a torus and a bilayer system with twist defects}

To begin, we consider topological ground state degeneracies on a 2-torus ($T^2$), which can be described by points on the plane with periodic boundary conditions along
two vectors $\hat{e}_1$ and $\hat{e}_2$, parameterized in the complex plane by $(1,\tau)$, as shown in Fig.~\ref{fig:setup}(a). We further define the two non-contractible cycles of the torus as $\alpha$ and $\beta$.

The topological ground state subspace (the code space) $\mathcal{H}_{T^2}$ is spanned by topological charge values assigned to a given non-contractible loop. 
For example, one can choose the $\alpha$-loop to label the basis states, such that $\ket{a}_\alpha=W^\alpha_a \ket{\mathbb{I}}_\alpha$ is created by inserting a Wilson loop
of an anyon $a$ along the $\alpha$-loop into the trivial vacuum sector $\ket{\mathbb{I}}_\alpha $. There are $n$ types of possible anyon charges, where $n$ is the ground-state degeneracy. These states are eigenstates of the Wilson loop 
$W^\beta_b$, which describes an anyon $b$ encircling the $\beta$-loop. For example, in the familiar case of the toric code ($\mathbb{Z}_2$ spin liquid) with 4-fold ground-state degeneracy \cite{kitaev2003}, such loop operators are 4 types of different strings of Pauli operators that  cross the system in two different directions, corresponding to the  4 types of anyon charges: $\mathbb{I}$ (vacuum), $e$ (spinon), $m$ (vison), and $em$ (fermion). Alternatively, one can choose the longitudinal $\beta$-basis, $\ket{a}_\beta = W^\beta_a \ket{\mathbb{I}}_\beta$. 
 
The mapping class group of a torus, MCG$(T^2)$=SL(2, $\mathbb{Z}$),  is generated by  $\displaystyle \mathcal{S}:\tau \rightarrow -\frac{1}{\tau}$
and $\mathcal{T}: \tau \rightarrow \tau + 1$. They correspond to the following transformations of the non-contractible cycles, $\mathcal{S}: (\alpha,\beta) \rightarrow (-\beta,\alpha)$ (exchanges the two cycles and flips the direction of $\beta$-cycle) and $\mathcal{T}: (\alpha,\beta) \rightarrow (\alpha+\beta,\beta)$ (a Dehn twist along the $\beta$-cycle), as illustrated in Fig.~\ref{fig:setup}(b, c). 
We also introduce the reflection operations which flip the direction along one of the two cycles, i.e., $\mathcal{R}_\alpha : (\alpha,\beta) \rightarrow (-\alpha, \beta)$ and $\mathcal{R}_\beta:(\alpha,\beta) \rightarrow (\alpha, -\beta)$. 

Now we review the representation of MCG$(T^2)$ and the corresponding unitary transformations in the Hilbert space $\mathcal{H}_{T^2}$ \cite{wang2008, Nayak:2008dp}. The modular $\mathcal{S}$ transformation \cite{footnote1} exchanges the loops [Fig.~\ref{fig:setup}(e)], i.e., $\mathcal{S} \ket{a}_\beta $$=$$\ket{a}_{\alpha}$ and $\mathcal{S} \ket{a}_\alpha$$=$$\ket{a}_{-\beta}$$=$$ \mathcal{R}_\beta \ket{a}_{\beta}$. Note that the $\mathcal{S}$ transformation can also be interpreted as a basis transformation between the two dual bases, i.e., $\ket{a}_{\alpha}$$=$$\sum_b \mathcal{S}_{ab} \ket{b}_\beta$, where $\mathcal{S}_{ab}$ is the modular $\mathcal{S}$ matrix. In the Heisenberg picture, the Wilson-loop operators are hence transformed as $\mathcal{S} W^\beta_a \mathcal{S}^\dagger$$=$$W^{\alpha}_a $ and $\mathcal{S} W^\alpha_a \mathcal{S}^\dagger $$=$$ W^{-\beta}_a$.   Similarly, the modular $\mathcal{T}$ transformation performs a Dehn twist on the Wilson loops [Fig.~\ref{fig:setup}(f)], i.e., $\mathcal{T} \ket{a}_\alpha$$=$$ \ket{a}_{\alpha+\beta}$ and $\mathcal{T} W^{\alpha}_a \mathcal{T}^\dag$$=$$W^{\alpha+\beta}_a $.  The modular $\mathcal{S}$ and $\mathcal{T}$ matrices can be represented in the ground state subspace 
in a particular basis (e.g.~$\ket{a}_\beta$) and by anyon diagrams as:
\includegraphics[width=0.8\columnwidth]{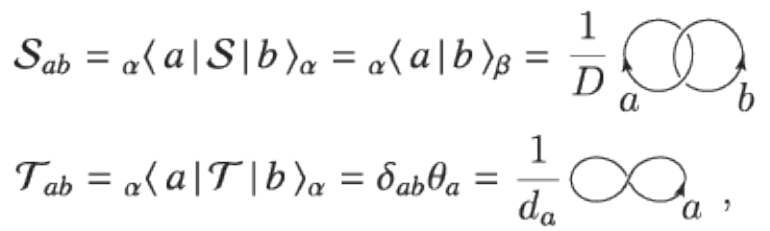}

\nin where $d_a$ is the quantum dimension associated with anyon $a$ and  $D$ is the total quantum dimension, $D=\sqrt{\sum_a d^2_a}$.  
As suggested by the Hopf-link anyon diagram, the $\mathcal{S}$ matrix encodes the self- and mutual-braiding statistics in its diagonal and off-diagonal elements respectively.  
On the other hand, the $\mathcal{T}$ matrix encodes the topological twist (exponentiated topological spin) $\theta_a$ in its diagonal elements.
As a simple example, the $\mathcal{S}$ and $\mathcal{T}$ matrices of the toric code ($\mathbb{Z}_2$ spin liquid) are shown below (in the basis formed by the 4 topological sectors $\ket{\mathbb{I}}_\beta$, $\ket{e}_\beta$, $\ket{m}_\beta$, and $\ket{em}_\beta$ ):
\be\label{modular_matrix_Z2}
\mathcal{S}_{Z_2}=\frac{1}{2}\begin{pmatrix}
    1&1&1&1\\
    1&1&-1&-1\\
    1&-1&1&-1\\
    1&-1&-1&1
\end{pmatrix},
\ \mathcal{T}_{Z_2}=\text{Diag}(1,1,1,-1),
\ee
with trivial self-braiding statistics and non-trivial mutual braiding statistics, and also nontrivial topological spin for the $em$ quasi-particle, i.e., $\boket{em}{\mathcal{T}}{em}=-1$.
The other simple example is the $\nu$$=$$1/k$ Laughlin FQH state, with the matrices given by $\mathcal{S}_{ab}$$=$$\frac{1}{\sqrt{k}}e^{i2\pi a b/k}$  and $\mathcal{T}_{ab}$$=$$\delta_{ab}e^{i2\pi a(a+k)/2}$, where both the self- and mutual-braiding statistics are non-trivial. 
These unitary matrices also represent fault-tolerant logical gates in the context of topological quantum computation, as will be discussed in Sec.~\ref{sec:logical_gate}.

Another way to create a space that is topologically equivalent to a high genus surface is to consider a bilayer topological state
in the presence of branch cuts that connect the two layers. The endpoints of these branch cuts are
twist defects, referred to as genons, which effectively increase the genus of the space by introducing non-contractible loops that intersect only once [see Fig.~\ref{fig:setup}(g-i)] \cite{Barkeshli:2010bo, Barkeshli:2012kw, Barkeshli:2013da, Barkeshli:2014by}.  From Fig.~\ref{fig:setup}(g) we see two different types of non-contractible Wilson loop operators:  
$W^\alpha_a$ (blue) passes through two branch cuts (line defects) and travels through the other layer, while $W^\beta_b$ (red) goes around a branch cut and always 
remains in the same layer; since these two loops cross only once, they are equivalent to the two non-contractible cycles of the effective torus in Fig.~\ref{fig:setup}(d).

\begin{figure}
\includegraphics[width=1\columnwidth]{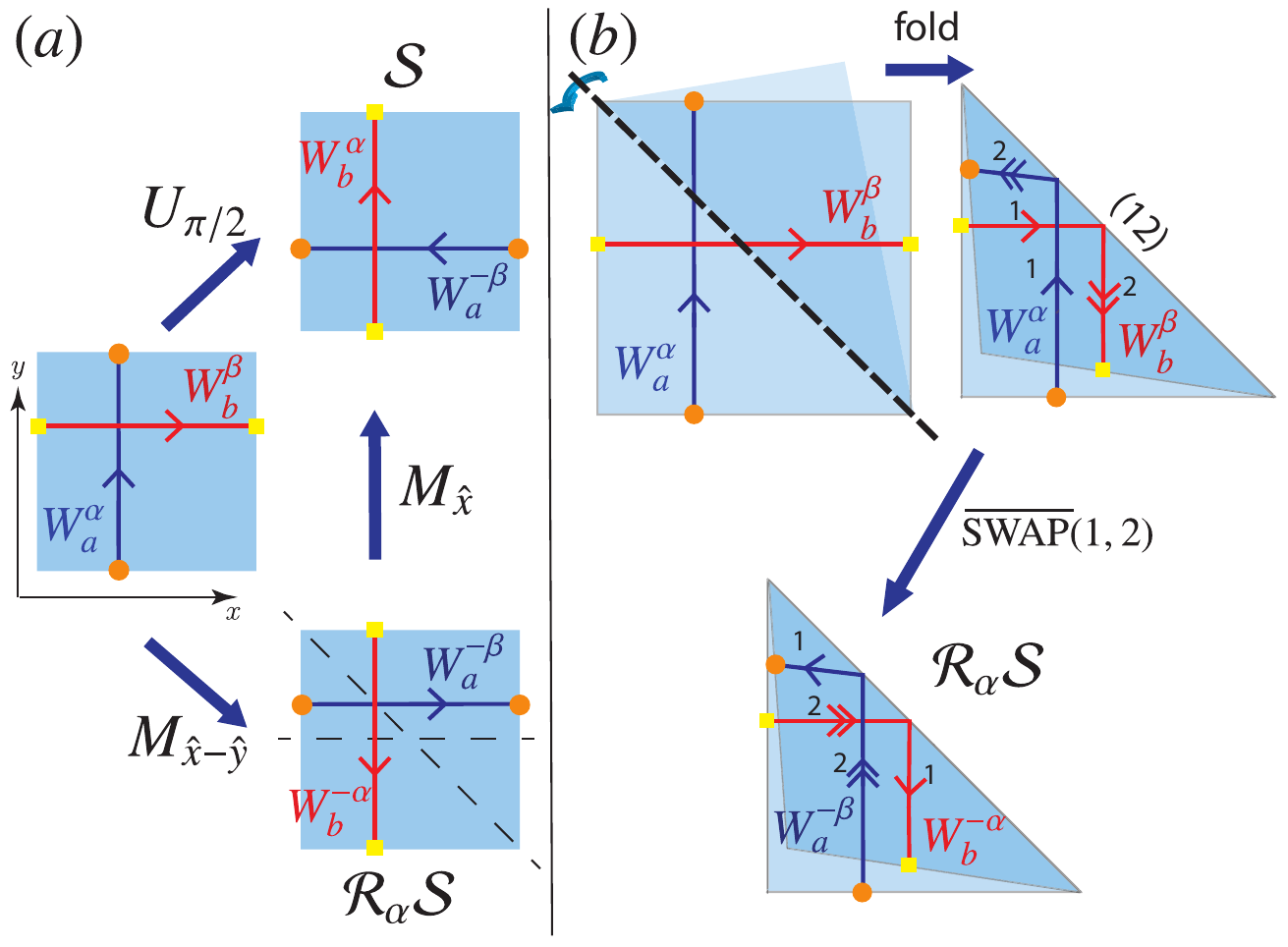}
\caption{(a) Modular $\mathcal{S}$ on a square with periodic boundary condition by $\pi/2$ rotation (equivalent to 
$M_{\hat{y}} M_{\hat{x} + \hat{y}}$), while $M_{\hat{x} + \hat{y}}$ alone implements $\mathcal{R}_{\alpha}\mathcal{S}$.  
(b) Folding the lattice into two layers along the diagonal line, where the left and bottom edges are still connected 
with periodic boundary condition. The upper-left edge has local boundary condition (12), connected with only 
local interaction. Different arrows represent loops in different layers as labelled. 
$\mathcal{R}_{\alpha} \mathcal{S}$ can be achieved by transversal SWAPs between layers 1 and 2.   }
\label{fig:triangle_protocol}
\end{figure}

\begin{figure*}
  \includegraphics[width=2\columnwidth]{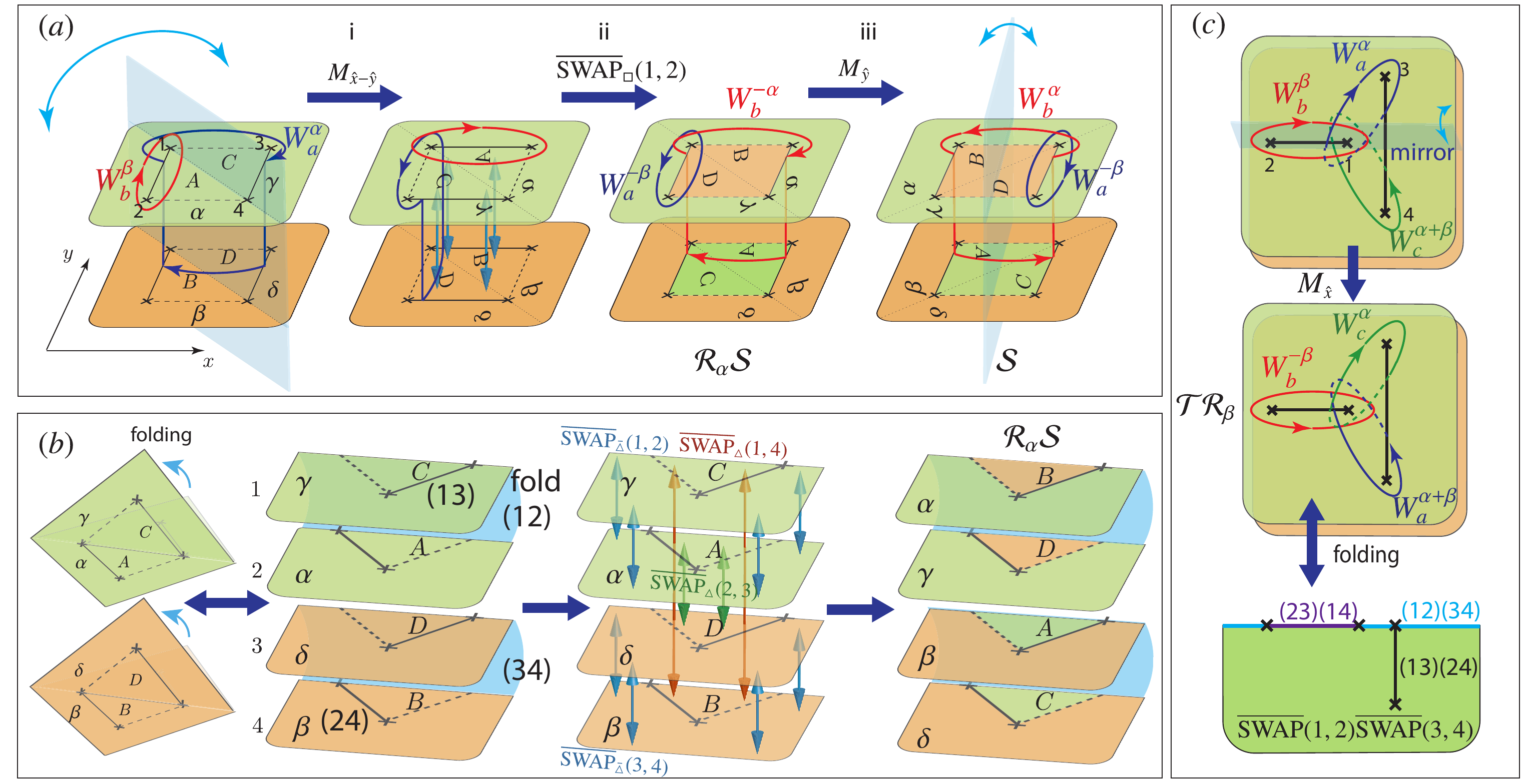}
  \caption{(a) Protocol for implementing $\mathcal{R}_{\alpha} \mathcal{S}$ and $\mathcal{S}$ in a bilayer topological state with genons.  The solid lines indicate twist defects (branch cuts); dashed lines
help illustrate the square patches. Note we assume the topological charge on the boundary is in the trivial vacuum sector, implying that the system 
can be treated as a sphere with two branch cuts, which is effectively equivalent to a closed torus. 
(b) Implementing $\mathcal{R}_{\alpha}\mathcal{S}$ on a the 4-layer folded system, with layer 1 and 2 (green) corresponding 
to the top layer of the original bilayer, and layer 3 and 4 (orange) to the bottom layer of the original bilayer. 
Layer 1 and 2 (3 and 4) are connected along the fold. The branch cuts connect layer 1 to 3, 
and layer 2 to 4.  $\mathcal{R}_{\alpha} \mathcal{S}$ 
can be implemented by a sequence of SWAPs between layers.  (c) A genon configuration with three-fold rotational symmetry. 
A single reflection $M_{\hat{x}}$ implements $\mathcal{T}\mathcal{R}_{\beta}$, equivalent to transversal SWAPs in the folded geometry. }
\label{fig:genon_protocol}
\end{figure*}

\section{Protocol for modular $\mathcal{S}$}

\nin Let us consider the case of a torus with $\tau =i$, corresponding to a square with opposite sides identified, so 
that the system has reflection and 4-fold rotational symmetries ($C_{4v}$ symmetry).  A $\pi/2$ rotation $U_{\pi/2}$ 
has the same effect as $\mathcal{S}$ on the non-trivial $\alpha$ and $\beta$ loops [Fig.~\ref{fig:triangle_protocol}(a)],
and therefore on the topological state space: $U_{\pi/2} |\psi \rangle = \mathcal{S} |\psi\rangle$ for $|\psi\rangle \in \mathcal{H}$ \cite{footnote2}. Alternatively, a mirror reflection ($M_{\hat{x} - \hat{y}}$) about the diagonal 
has the same effect as $\mathcal{R}_\alpha \mathcal{S}$ on the loops, and therefore on the topological state space: 
$M_{\hat{x} - \hat{y}} | \psi\rangle = \mathcal{R}_\alpha \mathcal{S} |\psi \rangle$. For a non-chiral reflection-symmetric topological state, $\mathcal{R}_\alpha$ 
acts entirely within the topological subspace $\mathcal{H}$; if the system is chiral then $\mathcal{R}_\alpha$ instead maps $\mathcal{H}$ 
to one associated with a reflected system. 

We begin with a toy example. By folding the square along the diagonal line as shown in Fig.~\ref{fig:triangle_protocol}(b), we 
obtain a 2-layer system defined on a triangle, where the right and bottom edges are connected by long-range couplings.  
The fold corresponds to a local boundary condition  $(12)$ with local interaction connecting layer $1$ and $2$ (written 
in cyclic notation of permutation group), which induces a gapped boundary.   In this geometry,  
the $\alpha$ and $\beta$ loops of the torus are related to each other by a permutation of the layers.  Thus 
we can consider the following protocol in the 2-layer system: apply $\overline{\text{SWAP}}(1,2) $$=$$ \prod_{j} \text{SWAP}_j(1,2)$, 
i.e., a product of pairwise local SWAP gates between sites (qubits) in layers $1$ and $2$ (enabled by inducing hopping for a certain time, see Appendix~\ref{append:SWAP_def}). This unitary transformation effectively implements $M_{\hat{x} - \hat{y}}$ on the torus. Therefore, $\mathcal{R}_\alpha \mathcal{S} |\psi \rangle $$=$$ \overline{\text{SWAP}}(1,2)  |\psi \rangle$, 
for $\ket{\psi}$ a topological ground state of the 2-layer system. As this operation acts on pairs of vertically separated 
qubits independently, i.e., on different subsystems under transversal partitioning, it is a transversal operation. For a topological phase of matter, these SWAP operations take a 
finite time to implement, and thus it is preferable to turn off the Hamiltonian before the protocol and turn it back on 
immediately afterwards; otherwise, the operation must be done on a time-scale much faster than that set by the energy 
scales of the Hamiltonian. In the context of active QEC, one applies a recovery operation by syndrome measurements to correct any local 
errors that are induced by inaccuracies in the SWAP operations (see Sec.~\ref{sec:error_fault-tolerant}).

We can also implement the $\pi/2$ rotation transversally by local SWAPs, generating $\mathcal{S}$, by folding the system three times.
This yields an 8-layer system with fully local boundary conditions, as discussed in Appendix \ref{append:8-layer}.  
In Appendix \ref{append:hexagon}, we further discuss a hexagonal geometry that allows implementation of $\mathcal{T}\mathcal{R}_\beta$ and $\mathcal{S}\mathcal{T}^{-1}$. 

The above protocols demonstrate a general concept: non-trivial modular transformations can be implemented as spatial
symmetry transformations when the system has rotation and/or reflection symmetries. By folding the system, and 
thus considering multiple copies of the system with appropriate boundary conditions, the spatial symmetry transformation 
can be converted to a spatially local layer permutation symmetry, which can then be implemented transversally by SWAP gates. 
While the unfolded system requires a certain point group symmetry (e.g. a square or a hexagonal lattice), the folded system
has no such constraints aside from a layer permutation symmetry, making it more widely applicable and versatile. 

However the above protocol has a major deficiency. The $\mathcal{R}_{\alpha} \mathcal{S}$ 
operation described in the folded two-layer system requires non-local interactions, as the left and bottom 
side of the triangle must be glued to each other. Moreover, the number of layers required to have local interactions is at least 6 (see Appendix \ref{append:hexagon}). 
A key result of this work is to demonstrate how to further apply this folding approach in a more general setting, 
to generate a larger group of modular transformations while requiring fully local boundary conditions and fewer layers. 
To do this, we need to instead consider generating a non-trivial surface using genons. 

In order to implement $\mathcal{S}$ and $\mathcal{R}_\alpha \mathcal{S}$ using genons, we consider the protocol listed below and illustrated in Fig.~\ref{fig:genon_protocol}(a):   
\textbf{i}.~$\text{M}_{\hat{x} - \hat{y}}\equiv \overline{\text{SWAP}}(x,y \leftrightarrow -y, -x)$:  apply mirror reflection along the diagonal axis $\hat{x} -\hat{y}$, 
equivalent to a long-range pairwise SWAP across the diagonal.  Now the locations of branch cuts change, and 
the state is thus a ground state of a different Hamiltonian, with a different orientation for the branch cuts.
To get back to the original topological subspace, apply \textbf{ii}.~$\overline{\text{SWAP}}_{\square}(1,2)$: transversal SWAP  between the top (green) and bottom (orange) square patches 
(surrounded by solid lines representing the branch cuts and two dashed lines), which leaves the central  patch with a different color.  
Steps i and ii achieve $\mathcal{R}_\alpha \mathcal{S}$. In order to implement $\mathcal{S}$, 
we also apply \textbf{iii}.~$M_{\hat{y}}=\overline{\text{SWAP}}(x,y \leftrightarrow -x, y)$. That is, apply another mirror operation along the $y$-axis. 

The above protocol uses long-range pairwise SWAPs in order to implement the mirror reflections. However if we consider folding the system,
then it is possible to implement the entire protocol with only local, vertical SWAPs between qubits in different layers. 
Specifically, for the $\mathcal{R}_\alpha \mathcal{S}$ protocol, one can fold the system to obtain a 4-layer system, as shown in Fig.~\ref{fig:genon_protocol}(b), 
where the fold is  (12)(34), i.e., connecting layer 1 and 2 (3 and 4).  The square region enclosed by the branch cuts then becomes a triangular
patch with edges defined by the gapped boundary (12)(34), and two types of branch cuts, which connect layers (13) and (24), respectively. 
In Fig.~\ref{fig:genon_protocol}(b), we translate the $\mathcal{R}_\alpha \mathcal{S}$ protocol to the folded system using only transversal SWAPs, 
\be
\mathcal{R}_\alpha \mathcal{S} =  \left[\overline{\text{SWAP}}_{\triangle}(1,4)(3,2)\right] \left[\overline{\text{SWAP}}_{\bar{\triangle}}(3,4)(1,2)\right],
\ee
when acting in the topological ground state subspace. Here $\triangle$ refers to the triangular patch (formed by branch cuts and dashed lines) 
in each layer and $\bar{\triangle}$ represents the region outside the triangular patch. Note that when allowing direct SWAPs between layers that are not nearest neighbors (such as layer 1 and 4), the above logical gate is exactly a depth-1 circuit. If such direct SWAPs are not allowed due to hardware constraint, that SWAPs between layers that are not nearest neighbors
can be reduced to a finite sequence of SWAPs between nearest neighbor layers. 

We further observe that $\mathcal{R}_\alpha$ can be implemented in the original bilayer setup (before the folding) by a reflection about the 
$\hat{y}$-axis, $M_{\hat{y}}$. This can be converted to a local pairwise SWAP transformation by folding the system about the $\hat{y}$-axis. 
Thus a 4-layer system can support transversal implementations of both $\mathcal{R}_\alpha$ and $\mathcal{R}_\alpha \mathcal{S}$, 
however each of these require a different configuration of defects and boundary conditions. To implement both operations transversally
in a 4-layer system then requires moving the defects to convert from one defect configuration to another.  It is also possible 
to directly implement $\mathcal{S}$  by a sequence of transversal SWAPs; this requires folding the system into 16 layers. 

By changing mirror axes, we can also implement $\mathcal{R}_\beta$ and $\mathcal{R}_\beta \mathcal{S}$. Moreover,
the topological charge conjugation operation $\mathcal{C} = \mathcal{S}^2$ can already be implemented 
transversally in the two-layer unfolded system of Fig. \ref{fig:genon_protocol}(a) by applying $\overline{\text{SWAP}}(1,2)$ (see Appendix \ref{append:loop-relation}).

\begin{table*}
\begin{tabular}{lllllllll}
\hline 
\hline  
Models 
 & \begin{tabular}{l}Logical \\ basis  \end{tabular}\phantom{spc} 
 &  \begin{tabular}{l}Modular $\mathcal{S}$\end{tabular}\phantom{spc}
 & \begin{tabular}{l}Modular $\mathcal{T}$\end{tabular} \phantom{spc} 
 & \begin{tabular}{l}Universality \\ of $\text{MCG}_{\Sigma}$\end{tabular} \phantom{spc}
\\
\hline 
Toric code   ($\mathbb{Z}_2$ spin liquid)  & $\ket{n_e n_m}_{\beta}$ (2-qubit; $n_e, n_m=0,1$) & $(\overline{H}_1 \otimes \overline{H}_2) \overline{\text{SWAP}}_{12}$ 
&Control-Z: $\overline{CZ}$ & Subgroup of Clifford  \\                            
\hline
$\nu=1/k$ FQH                            & $\ket{a}_{\beta}$ (qudit)                                       & Fourier transform            & Phase gate   & (Generalized) Clifford \\ 
        &  $a=0,1,.. k-1$  &   $\mathcal{S}_{ab}=\frac{1}{\sqrt{k}}e^{i2\pi a b/k}$  & $\mathcal{T}_{ab}=\delta_{ab}e^{i2\pi a(a+k)/2}$  \\  
\hline                                                                                                                                                    
 Double semion                 & $\ket{n_s}_{\beta}$ , (qubit, $n_s = 0,1$)                             & Hadamard $\overline{H}$             & Phase $\overline{P}$ & Clifford  \\
\hline  
Ising                                      &              $\ket{\mathbb{I}}_{\beta}$, $\ket{\sigma}_{\beta}$, $\ket{\psi}_{\beta}$                  & \multirow{3}{*}{$\mathcal{S} = \frac{1}{2}\left(\begin{matrix}
 1 & \sqrt{2} & 1  \\
 \sqrt{2} & 0 &-\sqrt{2} \\
 1 & -\sqrt{2} & 1  \\
\end{matrix}\right) $}   &  & Universal (+ topological \\ 
 &                                             (qutrit)                          &      & $\mathcal{T} = \text{diag} \left(1,  e^{i \pi/8}, -1 \right) $  &  charge measurements)  \\   
 & & &  \\        
\hline  
Fibonacci                          &       $\ket{\mathbb{I}}_{\beta}$, $\ket{\tau}_{\beta}$   (qubit)                               &    $\mathcal{S}=\frac{1}{\sqrt{2+\phi}} \left( ^{1 \ \ \phi}_{\phi \ -1} \right)$, $\phi=\frac{1+\sqrt{5}}{2}$
          & $\mathcal{T} = \left( ^{1 \ \ 0}_{0 \  e^{i 4\pi/5}} \right)$   & Universal    \\            
\hline  
\hline        
\end{tabular}
\caption{Examples of logical gates and universality of the gate set that can be achieved using
the mapping class group on a high genus surface $\Sigma$ for various well-known topological states. 
Logical basis on a torus is listed, using a subset of the torus states $|a \rangle_\beta$. 
The logical string operators of all the Abelian models are as follows. $\mathbb{Z}_2$ toric code: $W^{\alpha}_{e,m}=\overline{Z}_{1,2}$, 
$W^{\beta}_{m,e}=\overline{X}_{1,2}$, where $1$ and $2$ labels the logical qubits.  
$\nu = 1/k$ Laughlin FQH states: $W^{\alpha}=\overline{Z}$, $W^{\beta} = \overline{X}$. Double semion: $W^{\alpha}_s = \overline{Z}$, $W^{\beta}_s = \overline{X}$. 
Here, $\overline{X}$ and $\overline{Z}$ are (generalized) logical Pauli operators, $\overline{H}$ is logical Hadamard, $\overline{P}$ is the Clifford phase gate. 
We note that any non-chiral topological state can be realized as an error-correcting code by using their corresponding commuting projector 
Hamiltonians \cite{kitaev2003,levin2005,koenig2010,Bonesteel:2012fl}; all chiral states have a doubled non-chiral analog.}
\label{table:transversal_gate}
\end{table*}

\section{Protocol for modular $\mathcal{T}$}

\nin In order to implement a transversal gate that involves modular $\mathcal{T}$, one needs $C_{3v}$ symmetry 
(a central patch with the shape of a equilateral triangle), as shown in Fig.~\ref{fig:genon_protocol}(c). 
In particular, a mirror reflection along the horizontal axis acts on the loops as $\mathcal{T} \mathcal{R}_\beta $.
As in the previous discussion, we can then consider folding the system along the mirror axis, which yields
a 4-layer system, with two different types of gapped boundaries and a line defect extending into the bulk. The mirror reflection then becomes
a local SWAP operation between vertically separated layers, providing a transversal implementation of $\mathcal{T}\mathcal{R}_\beta $ with a depth-1 circuit, i.e.,
\be
\mathcal{T}\mathcal{R}_\beta =  \overline{\text{SWAP}}(1,2)(3,4). 
\ee 

In the same triangular geometry of the genons, $\mathcal{R}_\beta \mathcal{S}$ corresponds to 
a mirror reflection along the diagonal line connecting defects 1 and 4 [see Fig.~\ref{fig:genon_protocol}(c)], followed by a vertical SWAP
operation between the two layers. Folding along this diagonal mirror axis then gives a transversal implementation of 
$\mathcal{R}_\beta \mathcal{S}$ by local layer SWAPs in this geometry. 

Therefore the transformations, $\mathcal{R}_\beta \mathcal{S}$ and $\mathcal{T}\mathcal{R}_\beta $ can each be implemented transversally through local 
SWAP operations in a 4-layer system. However they each require different configurations of the line defects (due to having originated from different folds), 
and therefore cannot both be implemented transversally in the same geometry. One can show that by further folding the system, it is possible to arrive at a configuration of line defects in a 12-layer
system, which admits both $\mathcal{R}_\beta \mathcal{S}$ and $\mathcal{T} \mathcal{R}_\beta $ to be performed transversally (see Appendix \ref{append:12-layer}). 
Combining the two transformations, one can perform the modular transformation $\mathcal{T}\mathcal{S}= (\mathcal{T}\mathcal{R}_{\beta})(\mathcal{R}_{\beta} \mathcal{S})$. 
Since $\mathcal{C}$ can be implemented by $\overline{\text{SWAP}}(1,2)$ in the unfolded system, we can also obtain 
$\mathcal{R}_\alpha \mathcal{S}$ and $\mathcal{T}\mathcal{R}_{\alpha}$, etc.  

Above, we also observed that $\mathcal{R}_\beta$ can be implemented in a 4-layer system with an appropriate geometry of defects. We see that moving
the defect configuration between different transversal operations in the 4-layer system allows 
us to implement $\mathcal{T} = (\mathcal{T} \mathcal{R}_\beta) \mathcal{R}_\beta$, as well as the whole MCG$(T^2)$.

\vspace{0.1in}

\section{Fault-tolerant transversal gates}\label{sec:logical_gate}

\nin Due to the transversal nature of the SWAP operations, our scheme can be used as a fault-tolerant logical gate in a wide class of topological
codes, both abelian and non-abelian \cite{kitaev2003,levin2005,koenig2010,Bonesteel:2012fl}. 
Specifically, considering sites that are separated vertically as a single subsystem, we see that the SWAP gates are local gates acted within the same subsystem. These local gates do not have any coupling to other subsystems, and hence the 
error cannot be propagated to other susbsystems as well \cite{Eastin:2009cj, Beverland:2016bi} (see Sec.~\ref{sec:error_fault-tolerant}).
  
The modular $\mathcal{S}$, $\mathcal{T}$ and their combination with $\mathcal{R}_{\alpha,\beta}$ 
implement particular types of transversal logical gates depending on particular topological states, 
with examples given in Table \ref{table:transversal_gate} and more detailed illustration in Appendix \ref{append:fault-tolerant}.  A notable result is the circumvention of the no-go theorem given 
in Ref.~\cite{Beverland:2016bi}, which claims that the power of transversal gates decreases as the 
universality of representations of the MCG of the topological state increases. The circumvention becomes possible in our scheme
due to the non-trivial configurations of defects and boundary conditions, which violates the spatial homogeneity assumption 
of Ref.~\cite{Beverland:2016bi}. In particular, according to Ref.~\cite{Beverland:2016bi}, the Fibonacci code admits only 
trivial transversal gates in the homogenous case; as we have shown here, introducing defects allows us to perform 
at least two types of transversal gates on a single torus as shown in Table~\ref{table:transversal_gate}. 
A particularly important application is for the Ising phase, where the modular $\mathcal{T}$ transformation allows a 
single-qubit $\pi/8$ phase gate, which, in addition to the Clifford group that is generated by braiding of 
$\sigma$ particles and measurements, provides a universal gate set \cite{bravyi200universal,barkeshli2016mcg}.

\vspace{0.1in}

\begin{figure}
  \includegraphics[width=1\columnwidth]{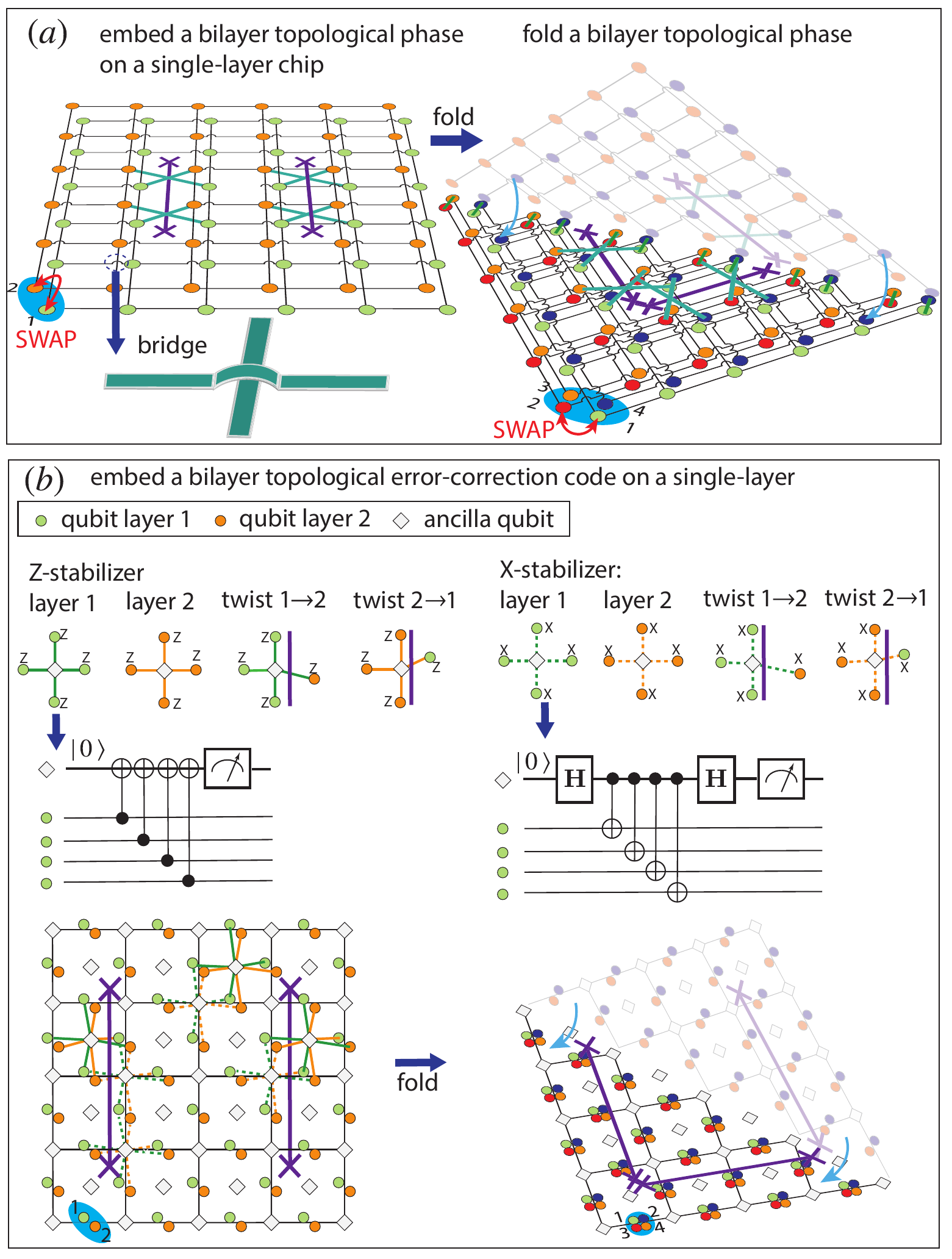}
  \caption{Experimental implementation with superconducting qubits: (a)  Embedding multi-layer topological phases in a single-layer chip, using crossover bridges. (b) Embedding multi-layer topological error-correction codes on a single-layer chip with a central ancilla performing stabilizer measurements.  When crossing a twist defect, the stabilizers connect qubits on different layers.}
  \label{fig:qubit_realization}
\end{figure}

\section{Possible experimental platforms} 

\nin We propose an experimental realization of the above ideas in superconducting qubit arrays, by considering a multi-layer version of topological phases \cite{Roushan:2016iu, Kapit:2014vba, Kapit:2015cy, Ma:2016hh,barkeshli2015} and error-correcting codes \cite{fowler2012,  Magesan:2015bg, Barends:2014fub, egrave:2015dx, Gambetta:2017iu, Bonesteel:2012fl, Wosnitzka:2016jt, Lekitsch:2015ua}. Multiple layers are embedded in a single-layer chip with a super-lattice structure, where each unit cell contains multiple sites, shown in different colors in Fig.~\ref{fig:qubit_realization}.  In the topological phase implementation, one can implement the nearest-neighbor  interaction with capacitive or inductive coupling between qubits, in the same sublattice, as shown in Fig.~\ref{fig:qubit_realization}(a).  Importantly, the crossover bridge architecture ensures that sub-lattices are not coupled to each other \cite{Chen:2014he, Foxen:2017te}. Different sublattices are only connected when a twist defect, shown as a purple line, is crossed. 
 
The active error-correction implementation is less demanding, since not all links are simultaneously connected \cite{fowler2012,  Magesan:2015bg, Barends:2014fub, egrave:2015dx, Gambetta:2017iu, Bonesteel:2012fl, Wosnitzka:2016jt, Lekitsch:2015ua}, and therefore, crossover bridges are not required.  Focusing on the toric code as the simplest example, shown in Fig.~\ref{fig:qubit_realization}(b), we start with a bilayer code embedded in a single-layer chip, highlighted by green and orange. A similar folded surface code architecture without twist defects has also been considered in Ref.~\cite{Moussa2016} in order to achieve the same set of transversal Clifford gates of a triangular color code Ref.~\cite{Bombin:2006hw, Kubica:2015br}. There are also some previous studies on adiabatic braiding  twist defects in the context of surface code quantum computation \cite{Brown2017}. 

Regardless of the layer, there are two types of stabilizer measurement gadgets, the ``plaquette'' ($Z^{\otimes 4}$) and ``star'' stabilizers ($X^{\otimes 4}$) \cite{fowler2012}, which can be implemented by dispersive coupling to the central ancilla resonator \cite{Nigg:2012wd}, shown as a white diamond. Each ancilla applies one of the stabilizers, and is connected to  8 neighboring qubits, with no crossing and requiring only two-qubit gates (CNOT) in between. To simultaneously implement the measurements on both layers, one can use an ancilla resonator with two modes, to independently couple to qubits of each layer. 
Alternatively we can use a single ancilla in two separate steps for each layer.  The stabilizers at the branch cuts connect qubits in two different layers, as illustrated in Fig.~\ref{fig:qubit_realization}(b). The measurement circuit for the non-abelian Fibonacci code can be found in Ref.~\cite{Bonesteel:2012fl}.

For both topological phases and codes,  the four-layer system can be obtained  by  ``code folding'' \cite{Nigg:2017}, 
which doubles the size of the unit cell. In this case, the transversal SWAPs, introduced above, become local nearest-neighbor SWAPs between qubits in the same cell.

\begin{figure}
  \includegraphics[width=1\columnwidth]{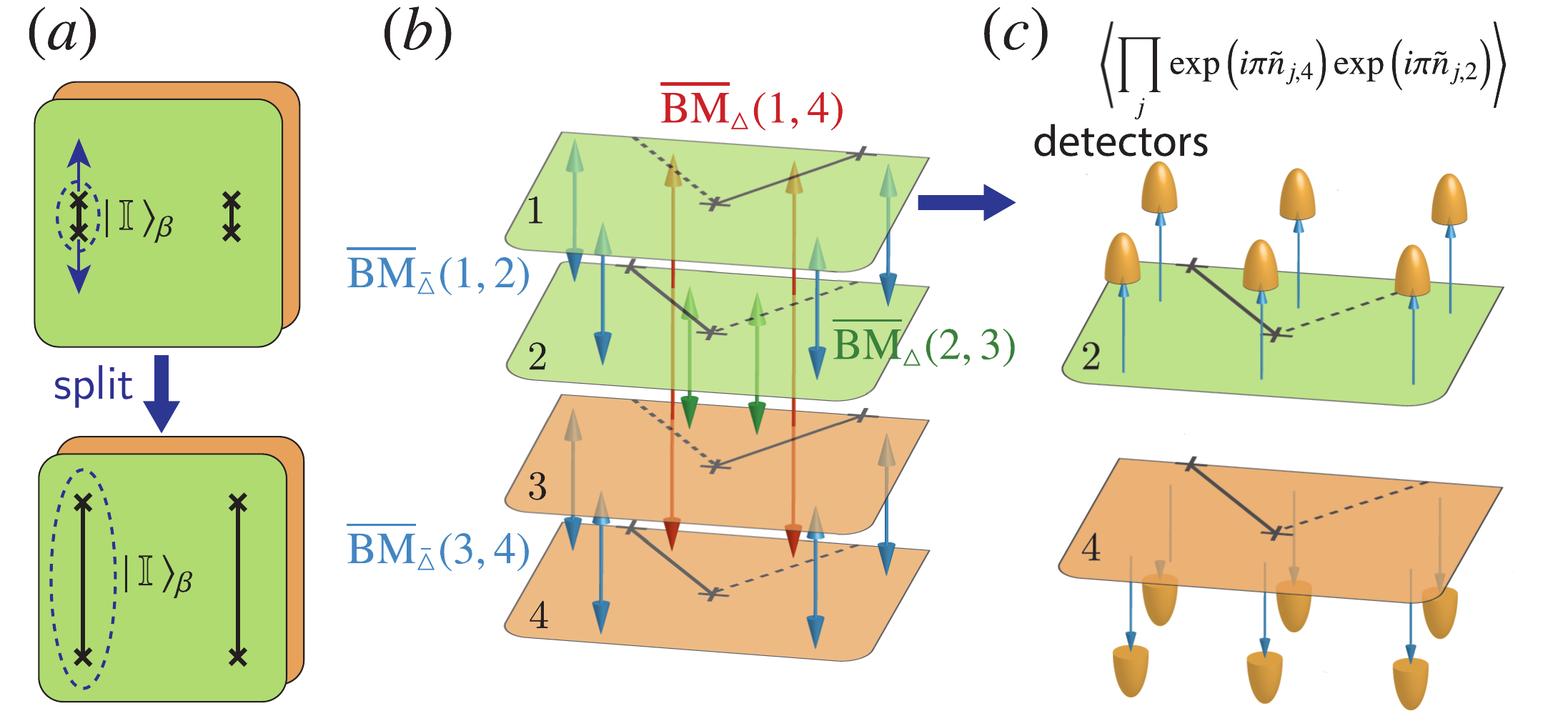}
  \caption{ Measurement protocol of modular matrices. 
  (a) Splitting two pairs of twist defects out of the ``vacuum", due to charge conservation, the ground state is in the vacuum sector $\ket{\mathbb{I}}_{\alpha}$ (trivial charge measured by the loop $W^{\beta}_a$ encircling the branch cut). 
  (b, c) Measuring the modular $\mathcal{S}$ matrix from local parity measurement after four parallel beam-splitter operations.}
\label{fig:beam_splitter}
\end{figure}
 
\vspace{0.1in}

\section{Interferometric measurement of modular matrices} 

\nin In this section, we show how to measure the modular matrices of a topological phase in order to 
diagnose its topological order and extract the braiding statistics and topological spins.  In particular, we consider two different approaches: the first one (in Sec.~\ref{sec:beam-splitter} and a generalized verion in Sec.~\ref{sec:measure_permutation}) uses beam-splitter operation and direct parity measurement, while the second one (Sec.~\ref{sec:Ramsey}) uses an extra ancilla to perform a Ramsey interferometry (also called Hadamard test) and quantum non-demolition (QND) measurement.

\subsection{Beam-splitter operation and parity measurement}\label{sec:beam-splitter}
For concreteness, we consider the case where we implement $\mathcal{R}_{\alpha}\mathcal{S}$.
We thus consider a four-layer state. 
The expectation value $\mathcal{R}_{\alpha}\mathcal{S}$ under a certain state $\ket{\psi}$ can be expressed as
$\boket{\psi}{\left[\overline{\text{SWAP}}_{\triangle}(1,4)(3,2)\right] \left[\overline{\text{SWAP}}_{\bar{\triangle}}(3,4)(1,2)\right]}{\psi},$
implying that one can measure it through the expectation value of a sequence of commuting SWAPs.

The first approach is using the many-body interferometer \cite{Daley:2012bd, Islam:2015cm} 
by applying a `beam splitter' (BM) operation between each pair of sites in two different layers, i.e., $\overline{\text{BM}}=\prod_j\text{BM}_j$, 
yielding the following mapping on the local operators in layer 1 and 2,
\be
\text{BM}_j: 
\left(\begin{matrix}
  a_{j,1} \\
  a_{j,2}  
\end{matrix}\right)
 \rightarrow
\left(\begin{matrix}
  \tilde{a}_{j,1} \\
  \tilde{a}_{j,2}  
\end{matrix}\right)
=\frac{1}{\sqrt{2}}\left(\begin{matrix}
  a_{j,1} + a_{j,2} \\
a_{j,1} - a_{j,2} 
\end{matrix}\right).
\ee 
 In this case, the global SWAP operator can be expressed and hence measured from the total parity of the particles in the anti-symmetric basis, i.e.,
$\left\langle \overline{\text{SWAP}}(1,2) \right\rangle$$=$$\left\langle \prod_j  \text{exp}\left(i \pi \tilde{n}_{j,2}\right)  \right\rangle, $ where $\tilde{n}_{j,2}=\tilde{a}^\dag_{j,2} \tilde{a}_{j,2}$ \cite{Daley:2012bd, Islam:2015cm, Pichler:2016ec},  (see Appendix \ref{append:SWAP_def}). 
Now,  as shown in Fig.~\ref{fig:beam_splitter}(c), we replace the four transversal SWAPs by four beam splitters operations: 
$\left[\overline{\text{BM}}_{\triangle}(1,4)(3,2)\right] \cdot\left[\overline{\text{BM}}_{\bar{\triangle}}(3,4)(1,2)\right]$,
followed by a parity measurement, yielding the expectation value
\begin{align}
\langle \mathcal{R}_{\alpha} \mathcal{S}\rangle =   \bigg\langle \prod_{j}  \text{exp}\left(i \pi \tilde{n}_{j,4}\right) \text{exp}\left(i \pi \tilde{n}_{j,2}\right)\bigg\rangle.
\end{align} 
One can use a similar protocol to measure $\langle \mathcal{T} \mathcal{R}_{\alpha}\rangle$ or $\langle\mathcal{S}\mathcal{T}\rangle$.

Below we assume a non-chiral system where all anyons are self-conjugate (such as $\mathbb{Z}_2$ spin liquid and double semion), 
for which $\mathcal{R}_\alpha \mathcal{S} = \mathcal{S}$. We first consider an equal mixture of all states in the 
ground state subspace, which corresponds to thermalizing the system at a temperature much below the gap ($T \ll E_g$), and
which can be represented by the density matrix $\rho=\frac{1}{N}\sum_a {_{\beta}}\ketbra{a}_{\beta}$, where $N$ is the ground-state degeneracy. 
The measurement of $\mathcal{S}$ under this ensemble thus becomes 
 \be
\langle \mathcal{S} \rangle= \text{Tr}\left(\frac{1}{N}\sum_a {_{\beta}}\ketbra{a}_{\beta} \mathcal{S}\right)=\frac{1}{N}\text{Tr}(\mathcal{S}).
 \ee  
The trace of the $\mathcal{S}$ matrix already contains useful information:  for example, for $\mathbb{Z}_2$ spin liquid $\text{Tr}(\mathcal{S}_{\mathbb{Z}_2})=2$, while the double semion phase has $\text{Tr}(\mathcal{S}_{D.S.})=0$.  Therefore, such a measurement could distinguish these two competing phases 
in the context of frustrated spin models.    

The diagonal modular matrix element can be obtained by the expectation value in a particular anyon sector $\ket{\psi}=\ket{a}_{\beta}$ 
[initialization and preparation discussed in Appendix \ref{sec:off-diagonal} and illustrated in Fig.~\ref{fig:beam_splitter}(a)].
 For example,   in the context of $\mathbb{Z}_2$ spin liquid [Eq.~\eqref{modular_matrix_Z2}], a
 nontrivial $\pi$-phase can be extracted from 
${_{\beta}}\boket{em}{\mathcal{T}\mathcal{R}_{\beta} }{em}_{\beta}={_{\beta}}\boket{em}{\mathcal{T}}{em}_{\beta}=-1$, 
originating from the topological spin of the fermionic spinon, labelled $em$, providing a key signature of the $\mathbb{Z}_2$ spin liquid.   Schemes for 
measuring off-diagonal elements are discussed in Appendix \ref{sec:off-diagonal}.

\subsection{Measurement of permutation operator}\label{sec:measure_permutation}

\nin Note that in the general case, the modular matrices are represented by a general permutation, and cannot be expressed as a product of commuting SWAPs (2-cycle), and hence measured by parity counting after the beam-splitter operation.  For example, to measure $\mathcal{S}$ for a chiral phase, one needs 8 layers, as shown in Appendix \ref{append:8-layer}, and to measure the operator: $\mathcal{S}$$=$$[\overline{\text{SWAP}}(1,8)(2,5)(3,6)(4,7)]\cdot[\overline{\text{SWAP}}(1,2)(3,4)(5,6)(7,8)]$ (applied from right to left). It is clear that $\overline{\text{SWAP}}(1,8)$ and $\overline{\text{SWAP}}(1,2)$ do not commute and cannot be measured simultaneously. Written in cyclic notation, one re-express the above permutation as $(18)(25)(36)(47)(12)(34)(56)(78)=(1735)(2648)$.  In general, one can always express an arbitrary permutation as a product of commuting cyclic permutations 
(N-cycle), as in this example. These cyclic permutations, also called twist operators $\overline{V}$, can hence be measured individually, 
as a generalization of the above SWAP protocol (for 2-cycle only). 

A local twist operator at site $j$, denoted as $V_j$,  applies a cyclic permutation $(012...N-1)$ on $N$ layers,  i.e.~
\be
V_j\ket{l}=\ket{l+1 \ \text{mod} \ N}, \ (l=0,1,2,\cdots N-1). 
\ee
It can be diagonalized in the Bloch eigenbasis as
\begin{align}
\non V_j \ket{\tilde{k}}_j = e^{i 2\pi k/ N} \ket{\tilde{k}}_j, \text{where } \ket{\tilde{k}}_j=&\frac{1}{\sqrt{N}}\sum_{l=0}^{N-1} e^{i 2\pi k l/N} \ket{l}_j, \\
 (k=&0,1,2,\cdots N-1).
\end{align}
Here $l$ is a layer index and $k$ labels the Bloch vector in the replica space.
In order to get the expectation value of the global twist operator, one can first perform a Fourier transform ($U_\text{FT}$) to the Bloch basis via beamsplitter operations, i.e.,
\be
\label{FTEq}
U_\text{FT}:a_{j,l} \rightarrow \tilde{a}_{j,k}=\sum_{k=0}^{N-1} e^{i2\pi lk/N} a_{j,l}
\ee
and count the number in each site $j$ of each layer $k$, labeled as $\tilde{n}_{j, k}=\tilde{a}^\dag_{j,k}\tilde{a}_{j,k}$,  and use the following formula \cite{Daley:2012bd} to calculate the expectation value:
\be\label{measure_twist}
\langle \overline{V} \rangle=\left\langle \prod_j V_j \right\rangle  = \left\langle \prod_j \prod_{k=0}^{N-1} \text{exp}\left(i \frac{2\pi}{N}  \tilde{n}_{j,k} k \right)  \right\rangle.
\ee
The whole protocol is reminiscent of the evaluation of Renyi entanglement entropy with replica trick \cite{Calabrese:2009dxa}. For the special case of $N=2$, the twist operator is the SWAP operator, and $k=0$ and $1$ corresponds to symmetric and anti-symmetric basis respectively.   

\begin{figure*}
\includegraphics[width=2\columnwidth]{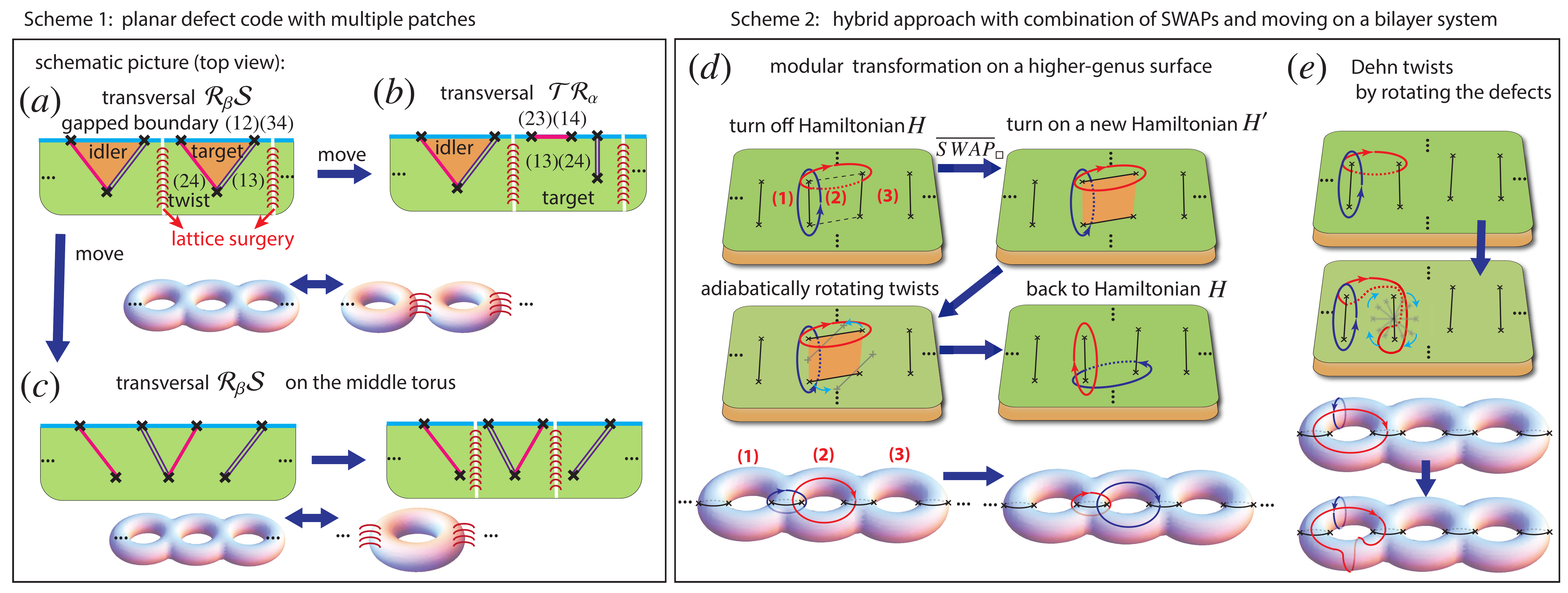}
\caption{Scalable schemes: 
   (a-c)  Planar defect code with multiple patches, where lattice surgery enables transversal gate within each patch.  (d-e)  Modular
   transformations on a higher-genus surface can be generated with the combination of SWAPs and moving of defects on a bilayer topological
   state. Branch cuts never have to be moved through each other, providing advantages over braiding-only schemes \cite{Barkeshli:2013da}.}
  \label{fig:scalable_architecture}
\end{figure*}

\subsection{Ramsey interferometer with an ancilla}\label{sec:Ramsey}

\nin A second approach to measuring the modular transformations is a QND measurement through a Ramsey-interference protocol (Hadamard test) with an extra ancilla qubit. The topological subspace is preserved after measuring the ancilla.

Below we will assume for concreteness, and without loss of generality, that the modular transformation being applied is $\mathcal{S}$. 
The Ramsey interference is performed by replacing all SWAP gates in the protocol with Control-SWAP gates, conditioned on an ancilla.
Measurement of the ancilla can then yield the wave function overlap between
the transformed and un-transformed states. 
The Control-SWAP operations can be realized with cold atoms by using a Rydberg atom as an ancilla \cite{Pichler:2016ec} or with cavity-QED interactions by using a cavity photon as an ancilla \cite{Jiang:2008gs, Zhu:2017vi} and is discussed in detailed in Appendix \ref{append:CSWAP}.

The interferometry protocol and quantum circuit are:
\includegraphics[width=1\columnwidth]{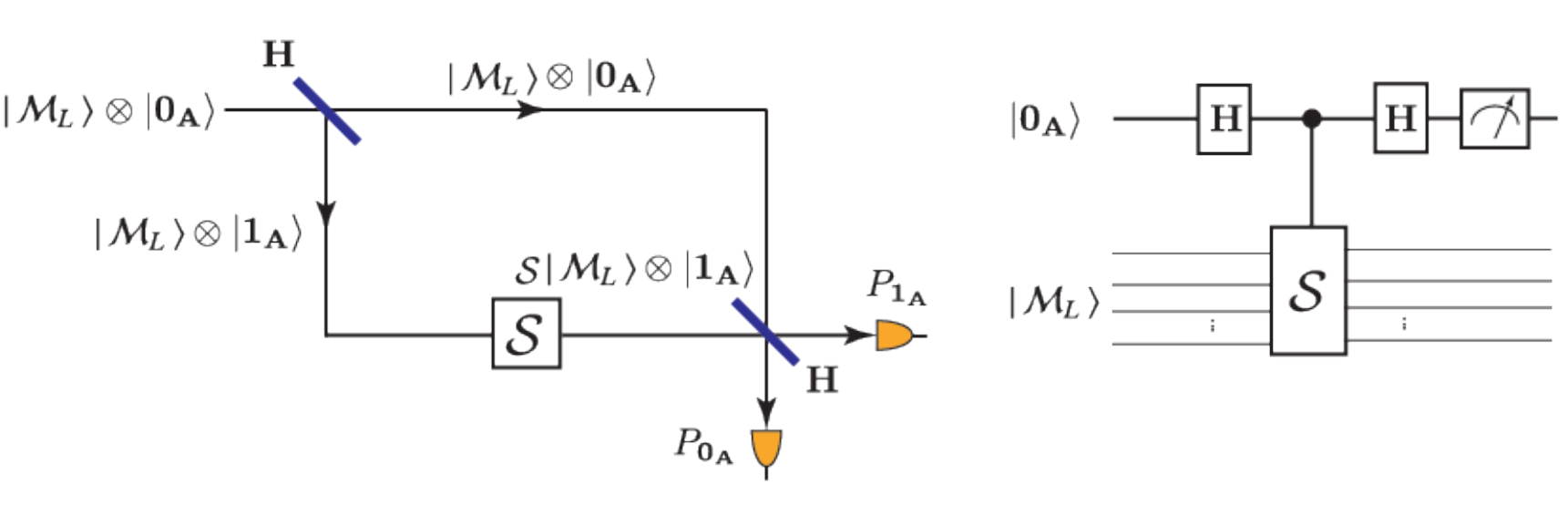}   
In the beginning of the protocol, one prepares the initial state as  $\frac{1}{\sqrt{2}}  \ket{\mathcal{M}_L}  $$\otimes$$  ( \ket{{\bf 0_A}} +   \ket{{\bf 1_A}} )$, where the ancilla
(denoted with a subscript ${\bf A}$) is in an equal-superposition state, and $\ket{\mathcal{M}_L}$ is the many-body topological state. 
The control-$\mathcal{S}$ operation then yields the state $\frac{1}{\sqrt{2}} ( \ket{\mathcal{M}_L}\otimes \ket{{\bf 0_A}} +  \ket{\mathcal{M}_R}  \otimes \ket{{\bf 1_A}} )$, 
where $ \ket{\mathcal{M}_R}$$=$$ \mathcal{S} \ket{\mathcal{M}_L}$.   By measuring the ancilla in either the Pauli $X$ or the $Y$ basis, one 
gets the real and imaginary parts of the overlaps:
\begin{align}
\non \langle X_{\bf A} \rangle =&\text{Re} [\bket{\mathcal{M}_L}{\mathcal{M}_R}] = \text{Re} [\boket{\mathcal{M}_L}{\mathcal{S}}{\mathcal{M}_L}], \\
\langle Y_{\bf A} \rangle =&\text{Im}[\bket{\mathcal{M}_L}{\mathcal{M}_R}] =\text{Im} [\boket{\mathcal{M}_L}{\mathcal{S}}{\mathcal{M}_L}].
\end{align}

\section{Scalable architecture}

\nin The preceding discussion can be generalized to genus $g$ surfaces by considering $n = 2(g+1)$ genons placed in a geometrically symmetric
configuration, for example at the vertices of a regular polygon with $n$ vertices (or $n-1$ vertices and one genon in the center). Then 
the rotational and reflection symmetries of the polygon implement non-trivial modular transformations. 
As above, folding the system in various patterns allows these transformations to be implemented by transversal SWAP operations
between layers. A crucial feature of folding is that microscopically any layer permutation is a possible symmetry
of a lattice Hamiltonian, whereas before folding, the possible rotational symmetries of a translationally invariant 
lattice Hamiltonian are limited. 

In the following, we briefly outline two additional scalable schemes for fault-tolerant quantum computation. 
The first scheme considers a plane composed of individual patches, where 
each patch effectively corresponds to a topological state on a single torus [see Fig.~\ref{fig:scalable_architecture}(a)]. 
One can use lattice surgery ideas \cite{horsman2012} to split the code into individual patches, perform the previously 
mentioned modular transformations transversally, and afterwards merge the code. Logical operations that entangle states 
associated to different patches (handles), could be achieved by adiabatically moving defects to the correct configuration 
in order to set up the required transversal gate. We leave a more detailed study of this for future work.

\begin{figure*}
\includegraphics[width=2\columnwidth]{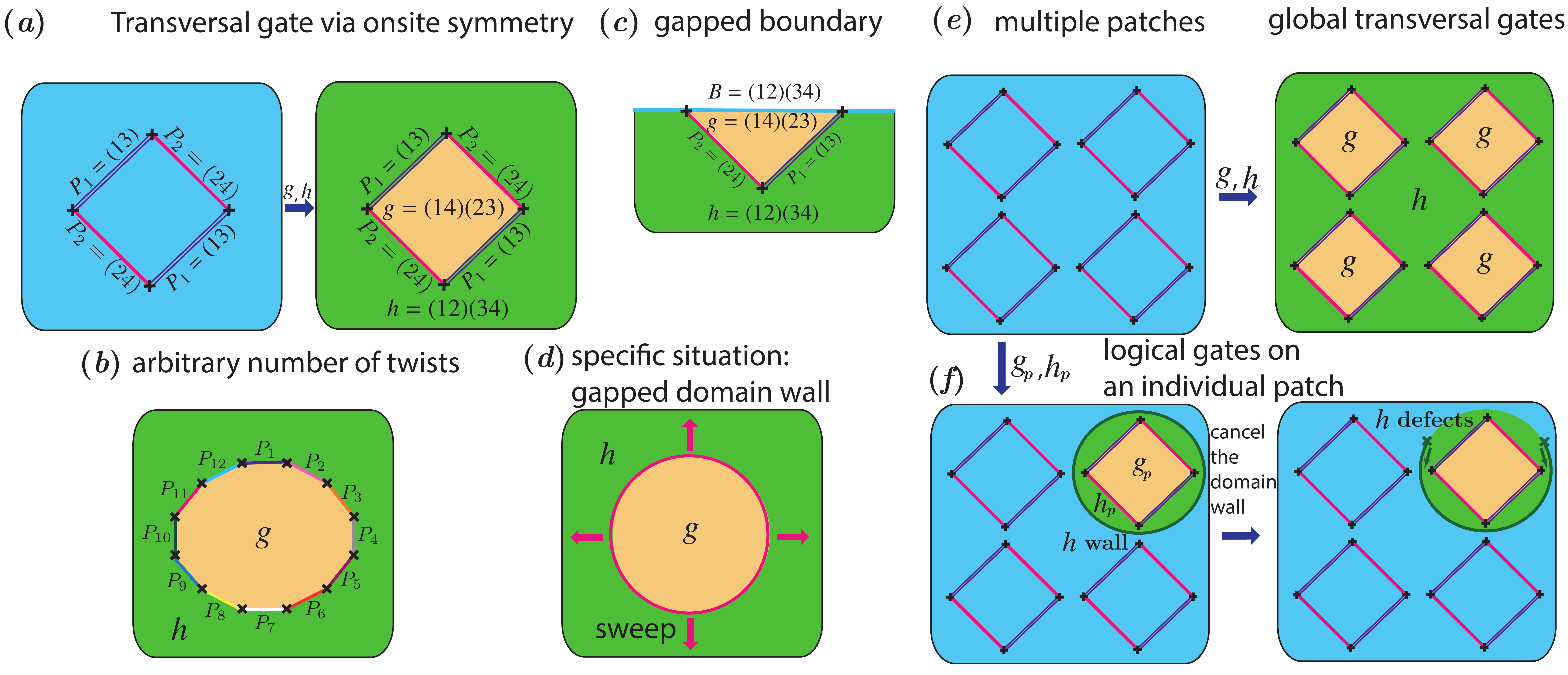}
\caption{Generalization to anyon symmetry transformations in the language of SET theory.   (a) Two types of twists ($P_1$ and $P_2$) encircle a patch. On the right panel, symmetry transformations $g$ and $h$ are applied inside and outside the patch respectively, as indicated by the change of colors. The transformation keep the twists invariant and hence preserve the code subspace.   (b) Generalization to a single patch formed by an arbitrary number of twists. (c) Generalization to a patch encircled by both twists and gapped boundaries. (d) A special limiting case with no twist defect and a single gapped domain wall circling the patch. Sweeping the wall across the code block leads to a transversal gate acting on the whole code block.  (e)  The more general situations with multiple patches composed by arbitrary number of twists.  The inside transformation $g$ and outside transformation $h$ lead to a global transversal gate acting on all logical qubits. (f) Apply logical gates on an individual patch and a subset of logical qubits.  First apply $g_p$ and $h_p$ on a region encircling the $p^\text{th}$ patch. A $h$ wall is automatically created by the $h_p$ and can be cancelled via creating a pair of $h$ defects, moving them around the loop and annihilate them again.}
  \label{fig:SET_illustration}
\end{figure*}

We can also consider a hybrid scheme, using a combination of vertical layer SWAPs 
and adiabatic moving of genons, without folding, as shown in Fig.~\ref{fig:scalable_architecture}(c-e). 
For the modular transformation in (d) resembling $\mathcal{S}$ on a single torus, we apply a SWAP between vertically separated sites from different
layers within a central square patch, and then adiabatically rotate the defects by $\pi/2$ to recover the original configuration of branch cuts. 
Depending on the application, the rotation could simply be stored in software, rather than physically implemented in hardware. 
A modular $\mathcal{T}$ (Dehn twist) on a single torus can be implemented by adiabatically rotating a branch cut by $\pi$, effectively braiding 
a pair of genons, as shown in (e) \cite{Barkeshli:2013da}. These operations can be used to implement a universal gate set for Fibonacci and Ising topological states (see Table~\ref{table:transversal_gate}). Compared to a scheme based on braiding defects alone \cite{Barkeshli:2013da}, in this hybrid
scheme we never have to take one branch cut through another. 

\section{Beyond modular transformations: generalization to generic anyon permutation symmetry and the relation to symmetry-enriched topological orders}

We note that although we have proposed experiments for physically folding the lattice into a multi-layer architecture, it is only necessary to do that in the context of measuring topological orders.  For topological quantum computation and error correcting codes, one does not necessarily need to have multiple physical layers and can instead always just use a single-layer code or lattice model containing an internal onsite symmetry that permutes anyon types. For example, the topological state could be described by two copies of anyon theories (mathematically known as modular tensor categories) $C \times C$, such that a $\mathbb{Z}_2$ on-site symmetry exchanges the anyons from the copies of $C$. That is, the anyon $(a,b)$, consisting of $a$ from one copy and $b$ from second copy, is transformed as $(a,b) \leftrightarrow (b,a)$.
A well-known example is the equivalence between the triangular color codes and a folded surface (toric) code, which both have a complete set of transversal Clifford gates \cite{Kubica:2015br}. There, the diagonal fold correspond to a specific gapped boundary $(e_1, e_2)(m_1, m_2)$ that allows exchange of the two types of $e$ and $m$ anyons in the two virtual layers. More generally, it has been shown how to construct lattice topological phases and codes with enriched on-site symmetries \cite{ChengSET2017, HeinrichSET2016}, known as symmetry-enriched topological states (SETs). In this regard, our ``quantum origami" scheme gives a general theoretical recipe for designing useful topological codes in the presence of on-site anyon-permuting symmetries together with twist defects and/or gapped boundaries, enabling powerful transversal logical gates.

Below we present a formalism to generalize the transversal gates discussed in this paper to a more general class of on-site symmetry transformations using the language of the SETs.   We consider a collection of twists/branch cuts (denoted by permutation group element $P_m$ for the $m^{\text{th}}$ twist) connected together to form a closed loop which divides a code into a  patch inside the closed loop of twists and a region outside the patch.  This is illustrated in Fig.~\ref{fig:SET_illustration}(a) for two  pairs of  twists, and the general case of arbitrary numbers of twists and encoded logical qubits in  Fig.~\ref{fig:SET_illustration}(c).  We can apply  an onsite symmetry transformation inside the patch, denoted by a group element $g$, and an onsite symmetry transformation $h$ outside the patch, such that $g$ and $h$ realize non-trivial permutations of the anyons.    
In order to ensure that the transformations form a logical gate $G$, we need to ensure that the transformation maps the code (ground) subspace $\mathcal{H}_\Sigma$ back to itself (i.e., forms an automorphism of $\mathcal{H}_\Sigma$),  which in turn requires that the twists ($P_m$) remain invariant under the symmetry transformations: $G: \mathcal{H}_\Sigma \rightarrow \mathcal{H}_\Sigma$.  This leads to the following condition:
\be\label{gate_condition}
g P_m h^{-1} = P_m \quad  (\text{for all $m$}).
\ee
The classification of all possible transformations $g$ and $h$ satisfying the above condition leads to a complete classification of all possible transversal gates in the corresponding twist configuration, which is left for future systematic development \cite{ZhuSET}.   As a simple limit of this condition, one can consider the twist being trivial ($P_m = I$). In this case, one can only have $g=h$, i.e., the same transformation inside and outside the patch, which obviously preserves the trivial twist and hence the code subspace $\mathcal{H}_\Sigma$.    Another simple example is illustrated in Fig.~\ref{fig:SET_illustration}(a) with 4 anyon labels (or 4 physical layers), similar to the construction for modular $R_\alpha \mathcal{S}$ in Fig.~\ref{fig:genon_protocol}(b) but without the gapped boundary (fold).  In this simple example, the two different types of  twists are represented by the permutation group elements as $P_1= (13)$ and $P_2= (24)$.  The symmetry  transformations inside and outside the patch are also represented by the permutation group elements: as $g=(14)(23)$ and $h=(12)(34)$. We can see that $gP_1h^{-1} = (14)(23)(13)(34)(12) = (13) =P_1$ and $gP_2h^{-1} = (14)(23)(24)(34)(12) = (24) =P_2$, which both satisfy the condition in Eq.~\eqref{gate_condition}, and hence form a transversal logical gate.  The symmetry transformation in the more general situation with arbitrary number of twists forming a patch (arbitrary number of encoded logical qubits) is illustrated in Fig.~\ref{fig:SET_illustration}(b).

Another generalization is to take into account gapped boundaries as shown in Fig.~\ref{fig:SET_illustration}(c), which has already been previously considered in our construction of modular transformation via folding in Fig.~\ref{fig:genon_protocol}(b).   In this situation, one should also make sure the transformation $g$ does not change the gapped boundary $B$, i.e., 
\be
gBg^{-1} = B, \quad   hBh^{-1} = B.
\ee
which gives an additional condition for the qualification as a transversal gate.   Indeed, in the  example shown in Fig.~\ref{fig:SET_illustration}(c), we have $gBg^{-1}= (14)(23)(12)(34)(23)(14) = (12)(34)= B$, similarly the other transformation for $h$ in the above condition also holds. 

A special limit of the general situations in Fig.~\ref{fig:SET_illustration}(a,b) is the limit that there is no twist defect and hence only a single gapped domain wall denoted by $P_1$, as illustrated in Fig.~\ref{fig:SET_illustration}(d). In this case, we can still satisfy the condition $g P_1 h^{-1} = P_1$ to form a transversal logical gate.  When sweeping the domain wall across the whole code block as shown in Fig.~\ref{fig:SET_illustration}(d), we effectively apply a transversal gate on the whole code block corresponding to the symmetry transformation $g$.  This specific case of the relation of TLG and the symmetry of a gapped domain wall has been previously discussed in Ref.~\cite{Yoshida2017, Yoshida2015, Webster2018} for Abelian topological stabilizer codes.   

Finally, our theory can be extended into an even more general situation with multiple patches, while each patch is formed by arbitrary number of twists, as illustrated in Fig.~\ref{fig:SET_illustration}(e) (for simplicity we only show 4 twists in each patch). Now we can apply a global transversal logical gate acting on each patch with an inside transformation $g$ and an outside transformation $h$ satisfying the condition in Eq.~\eqref{gate_condition}, as illustrated in Fig.~\ref{fig:SET_illustration}(e).  Alternatively, if we want to apply logical gates on a small subset of logical qubits, we need to be able to apply symmetry operations on an individual patch.   This can be achieved by acting a single inside transformation $g_p$ on the $p^\text{th}$ patch, and a single outside transformation $h_p$ on a region encircling the patch, satisfying $g_p P_m h^{-1}_p = P_m$, as illustrated in Fig.~\ref{fig:SET_illustration}(f).   After applying the transformations,  we will create a gapped domain wall on the outer boundary of the region where $h_p$ was applied.  We call this an $h$ wall, since it can be represented by the permutation group element corresponding to the $h_p$ transformation.   Interestingly, we can cancel the $h$-wall by creating a pair of corresponding twist defects, the $h$ defects, and moving them around the wall until they annihilate with each other again.   In this way, the combination of the transversal symmetry operation and moving of twist defects apply logical gates only on a subset of logical qubits.  This approach can also serve as an alternative to the lattice surgery approach proposed in Fig.~\ref{fig:scalable_architecture}(a-c), which now does not need any measurement since the entire protocol is unitary.  

In sum, our general theory uncovers a profound connection between symmetry-enriched topological (SET) states of  matter \cite{barkeshli2014SDG, ChengSET2017, HeinrichSET2016} and transversal logical gates, which leads to a nearly unexplored theoretical direction and anticipates further systematic development in future \cite{ZhuSET}.  Our study explicitly demonstrates how not only topological operations such as braiding, but also symmetry operations, can be used as fault-tolerant logical gates.

\section{Error and noise analysis}

\subsection{Error analysis of the fault-tolerant transversal gate}\label{sec:error_fault-tolerant}

\nin The modular transformations used here can be viewed as fault-tolerant logical gates for the purpose of quantum computation. 
In the language of  fault-tolerant quantum computation, our global SWAP gate between patches A and B, 
$\overline{\text{SWAP}}_{AB}$$ =$$ \prod_{j \in  AB} \text{SWAP}_j$ is transversal. That is, if we include qubits 
from different layers that are aligned vertically in a single subsystem $j$ under transversal partitioning,  then the global SWAP operation is a product of operators which do not couple different subsystems ($j$ and $j'$) in the same code block. Therefore the error within one subsystem ($j$) cannot be propagated into a different subsystem ($j'$). This implies that such gates are fault-tolerant, as we will  briefly elaborate below. 

Topological states can be used for fault-tolerant ``topological'' quantum computation in two distinct contexts, through
either an active or passive approach. In the active error correction approach, the topological state is the eigenspace of a 
set of commuting projectors, which can be continuously measured through the use of an ancilla and a local finite depth 
quantum circuit. In this approach the errors can be tracked through various ``decoding'' algorithms. One can then either
deal with the errors entirely in software, or apply a recovery procedure to remove them. 

In the passive approach to error protection, one considers a Hamiltonian whose ground state is the topological phase of interest, and cools
the system down to a temperature $T$ that is much less than the energy gap $\Delta$ above the ground state. The topological protection
in this case is due to the exponential prefactor $e^{- \Delta /T}$ for the density of quasiparticle excitations. 
 
Let us consider local coherent errors in the pairwise local SWAP operations. As discussed above, the local pairwise 
SWAP is generated by tunneling and an additional phase shift.   For simplicity, we express it with a single generator 
as mentioned above: $U_j(t)$$=$$ \text{exp}\left(i 2Jt a^\dag_{j,-}a_{j,-}\right) $,  where $a_{j,-}=\frac{1}{\sqrt{2}}(a_{j,A}-a_{j,B})$ is the anti-symmetric mode.  Here, we have assumed the tunneling time $t$ equal to the phase shift time, which should be $t= \pi/(2J)$ to obtain the exact SWAP operator, namely $\overline{\text{SWAP}}_{AB}=\prod_{j \in  AB}U_j(\frac{\pi}{2J})=\overline{U}(\frac{\pi}{2J})$.   
   
Let us first consider an overall systematic error in the tunneling time: $\overline{U}(\frac{\pi}{2J}+\Delta t) = \overline{U}(\Delta t)\overline{U}( \frac{\pi}{2J})$. 
We note the additional erroneous unitary $\overline{U}(\Delta t)$ is actually factorizable, i.e., $\overline{U}(\Delta t)=\prod_{j \in  AB} U_j(\Delta t)$, 
therefore the local error does not spread out in space. The original wavefunction $\ket{\psi_0}$ is in the topological ground state (code) subspace  $\mathcal{H}$.  
The wavefunction after the operation without any errors, $\ket{\psi}=\overline{U}(\frac{\pi}{2J})\ket{\psi_0} = \overline{\text{SWAP}}_{AB} \ket{\psi_0}$, 
is also in the same code space $\mathcal{H}$ (this discussion can be straightforwardly generalized to the case where the initial and final code subspaces are different). 
On the other hand, the erroneous wavefunction becomes 
$\ket{\psi'}$$=$$\overline{U}(\Delta t) \ket{\psi} = \prod_j U_j(\Delta t) \ket{\psi}$.  If $\Delta t$ is a small quantity, one can perform a Taylor expansion: 
  \begin{align}
  \ket{\psi'}=&\ket{\psi}+ i (2J\Delta t)\sum_{j\in AB} a^\dag_{j,-} a_{j,-} \ket{\psi}+ O(\Delta t^2). 
  \end{align}
   Note that the major component $\ket{\psi}$ is still within the code space $\mathcal{H}$, while the first and higher order corrections 
with extra tunneling are outside $\mathcal{H}$.   Due to the factorizability of the erroneous unitary, the leading low-order corrections 
all have only local errors, or in other words, local generation of anyons (quasiparticles). Non-local errors with weight larger than half of the code 
distance $d/2$ (causing a logical error) is of $O[(\Delta t)^{d/2}]$ and is hence exponentially suppressed. 

The above discussion straightforwardly generalizes to the case of local independent errors for each site, 
i.e.~$\overline{U}(\{\frac{\pi}{2J}+\Delta t_j\})=\prod_{j \in  AB} \overline{U}({\Delta t}_j)\overline{U}( \frac{\pi}{2J})$. 
The error is again factorizable, leading to a Taylor expansion:
 \be
  \ket{\psi'}=\prod_j U_j(\Delta t_j) \ket{\psi}=\ket{\psi}+ i (2J)\sum_{j\in A, B} \Delta t_j a^\dag_{j,-} a_{j,-} \ket{\psi}+ O(\Delta t^2).
 \ee

In the context of active error correction, all the terms with order lower than $O[(\Delta t)^{d/2}]$  can be corrected using standard error 
correction protocols, where errors are either removed by a recovery procedure \cite{Raussendorf:2012ht}, or simply kept track of entirely in software. 

In the context of passive protection by a Hamiltonian, we note that $\ket{\psi'}$ has virtual anyonic pairs which increases its energy by $4 J^2  \Delta t^2 E_g$, to leading order,
where $E_g$ is the energy gap (excitation energy of an anyon pair).  Any further separation of the anyon pair by a distance $l$
is an $l^\text{th}$-order process due to the presence of the energy gap, which is suppressed by a factor $(\frac{\kappa}{E_g})^l$, where $\kappa$ is the perturbation strength 
that leads to anyon separation. With $\kappa < E_g$, such a virtual process, leading to a logical error when $l>d/2$, is exponentially suppressed.
Therefore, although the wavefunction $\ket{\psi'}$ is distinct from $\ket{\psi}$ in terms of microscopic details, they are simply related to each other by local operators,
which does not affect the logical state upon which the Wilson loop operators act. 
 
We have seen above that an individual transversal logical gate is protected. However, the situation is more subtle when applying a sequence of 
transversal gates. After the application of $n$ such transversal gates, there is an accumulation of virtual anyon pairs and energy increase. After $n \sim O(d/2)$ steps, a logical error occurs with probability of $O(1)$.  Therefore, one needs to suppress the virtual anyon accumulation by coupling the system to a low-temperature 
bath, such that after the application of each transversal gate, the virtual anyonic excitation can be relaxed to the ground state through 
a dissipative process via the system-bath coupling \cite{Bombin:2013cv}, in analogy to the recovery operation in active QEC.   

We caution that, in 2D, a topological memory with a finite-temperature bath is not truly self-correcting in the sense that there is always a 
finite maximum lifetime for the logical quantum information that exponentially increases with the gap size and is independent of system size \cite{Brown:2016cla}.
The transversal gates are still protected within this life time. Nevertheless, for higher dimensions (e.g., in 4D), a true self-correcting memory can exist, where 
similar transversal SWAP gates between different sub-lattices  can also be applied.

\subsection{Error analysis of the measurement protocols}

\nin In the previous sections, we have proposed two methods for measuring modular transformations: (1) through a Ramsey interferometer (Sec.~\ref{sec:Ramsey}), which uses an ancilla, and 
(2) by directly measuring SWAP operators (Sec.~\ref{sec:beam-splitter}), as in a many-body interferometer with beam splitters. In this section we briefly describe the effect
of (a) errors in the SWAP operations, (b) deviations of the system from the pure ground state due to finite temperature or perturbations
to the Hamiltonian, and (c) errors in the measurement process. Below we briefly discuss these issues. 
 
The Ramsey interferometer essentially takes the overlap between the wavefunctions at the end of two different histories,  
$\ket{\psi_0}$ and $\mathcal{R}_{\alpha} \mathcal{S}\ket{\psi_0}$. Here we consider the simplest protocol,
which implements $\mathcal{R}_\alpha \mathcal{S}$, although the discussion can be readily generalized to other protocols as well. 
As discussed in the main text, the $\mathcal{R}_{\alpha} \mathcal{S}$ operation can be implemented by two SWAP operations,
$\prod_j \text{SWAP}_j(1,2) \text{SWAP}_j(3,4)$. Let us suppose that there is an error in these operations due to error in the
required tunneling time needed to implement the SWAP operations. The final wave function, after the controlled-SWAP, is then
$\ket{\psi'_f}= \frac{1}{\sqrt{2}}\big[\ket{M}_L \otimes \ket{{\bf 0_A}} + \mathcal{M}_R \otimes \ket{{\bf 1_A}}\big]$,
where $\ket{\mathcal{M}_L} = \ket{\psi_0}$ and $\ket{\mathcal{M}_R} = \prod_j U_{j}(\Delta t_j) \mathcal{R}_\alpha \mathcal{S} \ket{\psi_0} $.
Here $U_{j}(\Delta t_j) \approx 1 +i (2 J) \sum_j (\Delta t_j^{(12)} a_{j,12-}^\dagger a_{j,12-} + \Delta t_j^{(34)} a_{j,34-}^\dagger a_{j,34-})$. $\Delta t_j^{(ab)}$ are the timing errors
for the SWAP operation between layers $a$ and $b$ for site $j$. $a_{j,12-}$ and $a_{j,34-}$ are the anti-symmetric modes associated with layers 12, and 34, respectively. 
 
 We expand the error of the wavefunction in the right branch of the history up to first order, i.e.~
\be
\ket{\mathcal{M}_R}=c\bigg[1+i \ 2J\sum_{j}\sum_{x = 12, 34} \Delta t_j^{(x)} a^\dag_{j,x,-} a_{j,x,-}\bigg]\mathcal{R}_\alpha\mathcal{S} \ket{\psi_0} + O(\Delta t^2),
\ee
where $c \approx 1/\sqrt{1+ (2J\sum_{j,x} \Delta t_j^{(x)} )^2} \approx 1-2 N^2 J^2 \Delta  t^2$ is the normalization constant (for wavefunction up to first order in $\Delta t$).  
Here, $N$ is the number of sites in the patches, and $\Delta t = \frac{1}{N}\sum_{j,x} \Delta t_{j}^{(x)}$ is the average error.  

The unperturbed and perturbed parts of the wavefunction are in general orthogonal to each other due to the creation of anyons.   
Therefore, the wavefunction overlap $\bket{\mathcal{M}_L}{\mathcal{M}_R} \approx (1-2N^2 J^2\Delta t^2) \boket{\psi_0}{\mathcal{R}_\alpha \mathcal{S}}{\psi_0} $. The
timing error thus results in an overall amplitude decay for the matrix element of the modular transformation. Since it is an overall amplitude decay, this implies that ratios of the
matrix elements of the modular transformations are insensitive to these timing errors, assuming the timing errors are independent of the initial state. 
Furthermore, we note that the overall amplitude decay is a power-law form, not an exponential, assuming $\Delta t J\ll 1/N$. 
Therefore a system with $N \approx 50$ particles and the state-of-the-art two-qubit gate fidelity 
$F \approx 1- J \Delta t  \approx 99.5\%$\cite{Barends:2014fub} would lead to a $12.5\%$ reduction 
of the wavefunction overlap.

Another source of error occurs due to a finite thermal density of quasiparticles at finite temperature $T$. These are suppressed by a factor of $e^{E_g/ T}$ where $E_g$
is the energy gap, at low temperatures, and this implies a small reduction in the expectation value. Assuming that $E_g$ corresponds to creating
two quasiparticles out of the vacuum, we expect that to leading order, the fidelity is reduced roughly by a factor of $\frac{1}{(1 + N^2 e^{-E_g /T})} \approx 1 - N^2 e^{-E_g/T}$, 
where the weight $N^2$ originates from the number of possible configurations of two quasiparticles. 

One can consider the case of additional disorder or perturbations to the Hamiltonian which makes the two patches (A and B) that are being swapped non-identical
microscopically. If the perturbation is small compared to the energy gap, the wave function again has perturbations that lead to an amplitude reduction 
set by a term that scales as $N^2$, as in the above examples. 
 
Finally, for the case of the many-body interferometer without an ancilla (Sec.~\ref{sec:beam-splitter}), there is a an extra error due to the final parity measurement. Specifically, since one is directly measuring the 
global parity operator ${\prod_{j}  \text{exp}\left(i \pi \tilde{n}_{j,4}\right) \text{exp}\left(i \pi \tilde{n}_{j,2}\right)}$ as mentioned above, the measured value is a product of all the local parity measurements. Thus, the measurement fidelity has an extra exponential decay, $F^N$, with $F$ being the average readout 
fidelity on each site.  The simplicity of the measurement scheme without a global ancilla comes at this price of exponential readout sensitivity.
This is the same challenge in the current entanglement entropy measurements in cold atom systems \cite{Islam:2015cm}.  
From a practical point of view, this parity measurement scheme is more suitable for small lattices with high readout fidelity. 
Examples with state-of-the-art parameters: (1) Superconducting qubits \cite{Barends:2014fub} $(99\%)^{50}=60.5\%, (99\%)^{30}=74.0\%$. (2) Ion qubits \cite{Harty:2014cm} $(99.93\%)^{100}=93.2\%,  \ (99.93\%)^{50}=96.6\%$. We caution that although the ancilla-based Ramsey interferometry 
does not suffer from the exponential readout sensitivity \cite{Pichler:2016ec}, the global ancilla may introduce correlated noise 
between distant sites, further complicating the fault-tolerance of the measurement.

\section{Conclusion and outlook}

\nin We have shown that when defects in a topologically ordered state are placed in a symmetric configuration, then spatial symmetry
transformations in a fully local system can effectively be identified with modular transformations. By folding the system, these spatial symmetry
operations become on-site symmetry transformations, allowing a simple method to implement modular transformations as a transversal
operation. This yields a novel method for directly measuring fractional statistics and also for implementing a wide class of transversal
logical gates. A even general theoretical framework about transversal gates as onsite anyon symmetry in the language of symmetry-enriched topological orders has also been presented, and anticipates further systematic classification in future \cite{ZhuSET}.

Extending the experimental implementation of these ideas to topological phases in multi-layer ultracold atomic systems in the future
allows the possibility of directly imaging topological order and fractional statistics. This can be done using local beam splitter
operations, as were successfully utilized in quantum gas microscopes for extracting entanglement entropy \cite{Islam:2015cm}. 

\vspace{0.1in}

\vspace{0.1in}
\nin \textbf{Acknowledgments}

We thank Michael Freedman, John Preskill, Zhenghan Wang, Sergey Bravyi, Steve Simon, Jeongwan Haah, and Ignacio Cirac for helpful discussions.     GZ and MH were supported by ARO-MURI, YIP-ONR, and the Sloan Foundation.  MB is supported by startup funds from UMD. The work by GZ was performed in part at Aspen Center for Physics, which is supported by National Science Foundation grant PHY-1607611.  All the authors were supported by NSF Physics Frontier Center at the Joint Quantum Institute.

\begin{appendix}

\section{Mapping class group of a torus and fractional statistics}\label{append:MCG} 

\nin A torus can be specified by points $z$ in the complex plane, modulo equivalences
$z \sim z + \omega_1 \sim z + \omega_2$, for complex numbers $\omega_1$ and $\omega_2$. 
The modular parameter is defined to be $\tau = \omega_1/\omega_2$. 

Arbitrary modular transformations belonging to the mapping class group (MCG) of a torus 
can be achieved by the following transformation
\be
\left(\begin{matrix}
  \omega_1 \\
  \omega_2  
\end{matrix}\right) \rightarrow
\left(\begin{matrix}
  \omega'_1 \\
  \omega'_2  
\end{matrix}\right)=
\left(\begin{matrix}
 a & b \\
 c & d 
\end{matrix}\right)
\left(\begin{matrix}
  \omega_1 \\
  \omega_2  
\end{matrix}\right), \quad \ \tau \rightarrow\tau'=\frac{a\tau + b}{c\tau + d},
\ee
satisfying $a,b,c,d\in \mathbb{Z}$ and $ad-bc=1$. Therefore, the mapping class group of a torus is 
isomorphic to a special linear group, namely MCG$(T^2)$=SL(2, $\mathbb{Z}$).  Note that the above transformation matrix $\left(\begin{matrix}
 a & b \\
 c & d 
\end{matrix}\right)$ is defined as a passive basis transformation.  An active transformation matrix on the loops is represented as its transpose, i.e., $\left(\begin{matrix}
 a & c \\
 b & d 
\end{matrix}\right)$, acting on loops such as those corresponding to the two cycles of the torus represented as $\alpha=\left(\begin{matrix}
 1   \\
 0  
\end{matrix}\right)$ and $\beta=\left(\begin{matrix}
 0   \\
 1  
\end{matrix}\right)$.
The SL(2, $\mathbb{Z})$ 
group has the following two generators represented by the active transformation matrix:

\nin (1) $\mathcal{S}=\left(\begin{matrix}
 0 & 1 \\
 -1 & 0 
\end{matrix}\right)$, giving the mapping $\displaystyle \mathcal{S}:\tau \rightarrow -\frac{1}{\tau}$.
This exchanges the two cycles of the torus and takes $(\alpha,\beta) \rightarrow (-\beta, \alpha)$. In the special case 
that the torus has a rectangular geometry, $\tau$ and $-1/\tau$ are pure imaginary (on the y-axis), 
therefore modular $\mathcal{S}$ is equivalent to a $\pi/2$-rotation.

\nin (2)  $\mathcal{T}=\left(\begin{matrix}
 1 & 0 \\
 1 & 1 
\end{matrix}\right)$, giving the mapping $\displaystyle \mathcal{T}:\tau \rightarrow \tau + 1$. 
Using our conventions for $\alpha$ and $\beta$, $\mathcal{T}$ thus corresponds to a Dehn twist
around the $\alpha$ loop, which takes the loops $(\alpha,\beta) \rightarrow (\alpha+\beta, \beta)$. 

Note that the $\mathcal{S}$ and $\mathcal{T}$ matrices defined above are associated with the mapping class group elements rather than their representation in the ground-state subspaces as mentioned in the main text [e.g.~Eq.~\eqref{modular_matrix_Z2} and Table \ref{table:transversal_gate}]. In this paper we also consider orientation-reversing maps:
$\mathcal{R}_{\alpha}=\left(\begin{matrix}
 -1 & 0 \\
 0 & 1 
\end{matrix}\right)$ and $\mathcal{R}_{\beta} = \left(\begin{matrix}
 1 & 0 \\
 0 & -1 
\end{matrix}\right)$. $\mathcal{R}_\alpha$ and $\mathcal{R}_\beta$ flip the winding numbers along the $\alpha$ and $\beta$ cycles,
respectively. In this representation, the charge conjugation operator flips the direction of the loops, and
corresponds to
$\mathcal{C} = \mathcal{R}_\alpha \mathcal{R}_\beta = \left(\begin{matrix}
 -1 & 0 \\
 0 & -1 
\end{matrix}\right)$. 

The transformations discussed in the main text also involve:
 $\mathcal{R}_{\alpha} \mathcal{S} = \left(\begin{matrix}
 0 & -1 \\
 -1 & 0 
\end{matrix}\right)$, $\mathcal{R}_\beta \mathcal{S} = \mathcal{C} \mathcal{R}_\alpha \mathcal{S} = \left(\begin{matrix}
 0 & 1 \\
 1 & 0 
\end{matrix}\right)$,  $\mathcal{T} \mathcal{R}_{\alpha} $$=$$\left(\begin{matrix}
 -1 &0 \\
 -1 & 1 
\end{matrix}\right)$ and $\mathcal{T} \mathcal{R}_{\beta} $$=$$\left(\begin{matrix}
 1 & 0 \\
 1 & -1 
\end{matrix}\right)$.  The matrices involving an odd number of reflections have Det$=-1$, and are hence orientation-reversing maps. 
Including these maps yields the extended mapping class group $\text{SL}^{\pm}(2, \mathbb{Z})$, which satisfies $ad-bc=\pm 1$.

\section{Definition and implementation of SWAP operators}\label{append:SWAP_def}

\nin The transversal SWAP between two vertically aligned patches A and B is defined as $\overline{\text{SWAP}}_{AB}$$ =$$ \prod_{j \in  AB} \text{SWAP}_j$, 
where the local pairwise SWAP operation has the property:
\be
\text{SWAP}_j \ket{\psi_j}_A \otimes \ket{\phi_j}_B = \ket{\phi_j}_A \otimes \ket{\psi_j}_B.
\ee
Here, $\ket{\psi_j}$ and $\ket{\phi_j}$ represent any arbitrary wavefunction on site $j$ belonging to  patches A and B, 
and their locations are switched by $\text{SWAP}_j$.   We consider an interlayer tunneling Hamiltonian 
\be\label{tunneling}
H_\text{t}= -J \sum_{j\in AB} (a^\dag_{j, A} a_{j, B}+ \text{H.c.}),
\ee
where $a_{j,A}$ represents the bosonic operator on patch A and site $j$. Physical implementation of SWAP can be achieved by  turning on this Hamiltonian for time $t=\pi/(2J)$, i.e.,  $\overline{U}^{(1)}(t)=e^{-i H_\text{t}t}$, with an additional phase shift $\overline{U}^{(2)}(t)$$=$$\prod_{j\in AB} e^{iJt(a^\dag_{j,A} a_{j,A} + a^\dag_{j,B} a_{j,B})}$,
i.e.,
\begin{align}\label{SWAPdefinition}
\non \overline{\text{SWAP}} =&\overline{U}\left(\frac{\pi}{2J}\right)= \overline{U}^{(1)}\left(\frac{\pi}{2J}\right)\overline{U}^{(2)}\left(\frac{\pi}{2J}\right) \\
=& \prod_{j\in  AB} e^{-i \frac{\pi}{2} (a^\dag_{j,A} a_{j, B} + \text{H.c.})}e^{i\frac{\pi}{2}(a^\dag_{j,A}a_{j,A} +a^\dag_{j,B}a_{j,B})}.
\end{align}

Now, in order to also measure the SWAP operator, we use the fact that it can be written as the parity operator 
in a rotated basis, after the application of the tunneling for time $t=\frac{\pi}{4J}$, up to a local phase. More specifically, 
we define the beam-splitter operator, $\overline{\text{BM}}=\overline{U}^{(1)}\left(\frac{\pi}{4J}\right)\overline{U}^{(2)}\left(\frac{\pi}{4J}\right)$.  
As mentioned in the main text, the beam splitter operation maps the operators in the two layers into the symmetric and anti-symmetric basis,
respectively. Note the SWAP operator in Eq.~\eqref{SWAPdefinition} can be re-written as the parity operator in the anti-symmetric mode of the two layers, i.e.,  
\begin{eqnarray}
\overline{\text{SWAP}} = \prod_{j\in AB} e^{i \pi \tilde{a}^\dag_{j,B}\tilde{a}_{j,B}},
\end{eqnarray}
 and can hence be measured after beam-splitter operations, where we have defined $\tilde{a}_{j,B}=\frac{1}{\sqrt{2}}(a_{j,A} - a_{j,B})$.
 Such a transversal SWAP operation and its  measurement, for example, has been recently achieved in cold atom experiments for the measurement of entanglement entropy \cite{Islam:2015cm}.

\begin{figure*}
\includegraphics[width=1.6  \columnwidth]{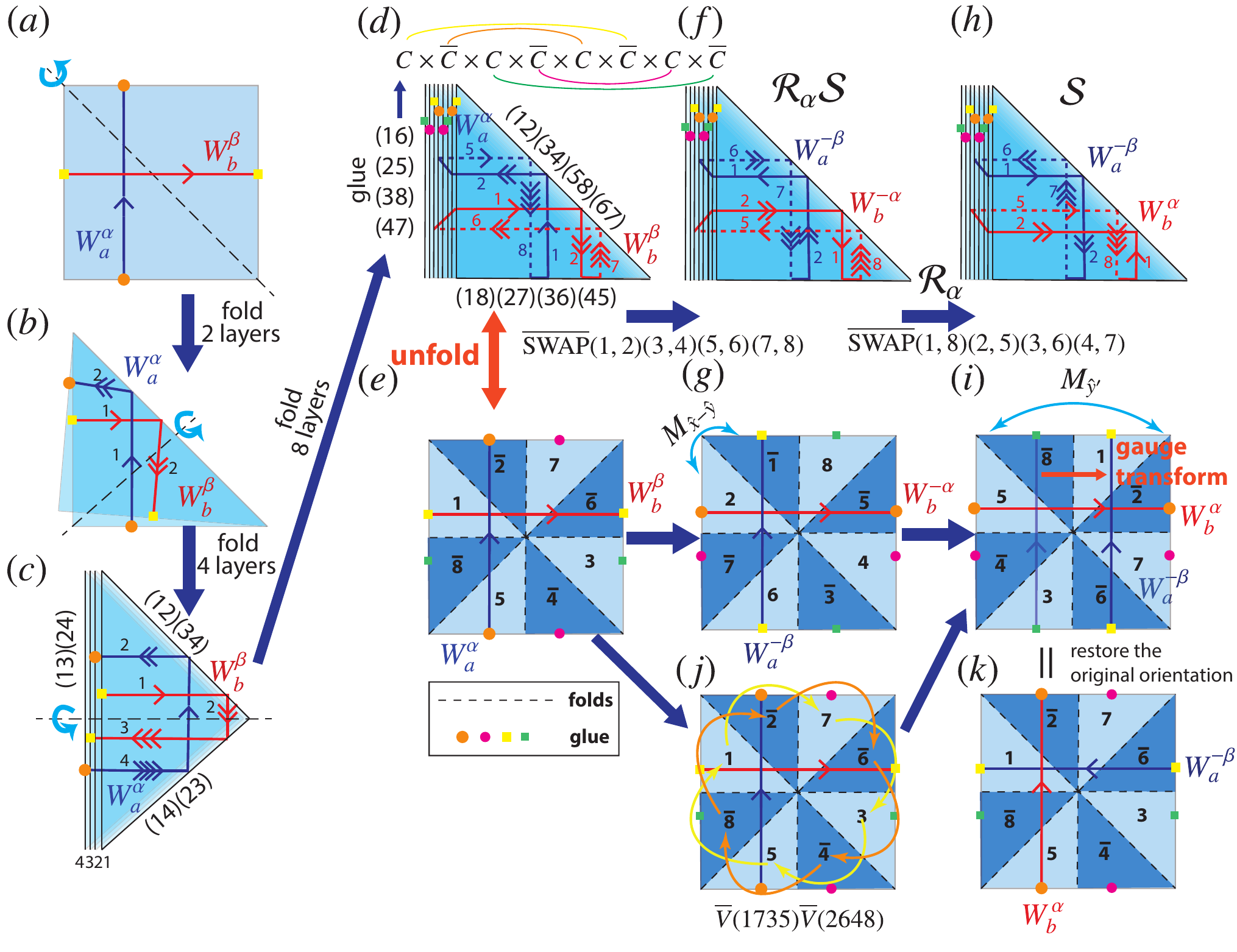}
\caption{(a-d) Folding the square with periodic boundary conditions three times lead to an 8-layer system with local boundary conditions induced by local interactions.
In panel (d), all the long-range boundary conditions come to the left edge and the layers with opposite chirality are glued locally.
This provides a novel way to effectively construct a torus in a planar system with local interactions. 
 (e) The correspondence to an unfolded system is shown by labeling the patches with the layer number. The bar on the number along with the dark-blue shade indicates layers with reversed chirality (orientation) in the folded system, compared to the original chirality. (f,g) The $\mathcal{R}_\alpha \mathcal{S}$ transformation can be implemented with four transversal SWAPs, corresponding to the mirror reflection along the diagonal line in the unfolded system.   (h, i)  The modular $\mathcal{S}$ transformation can be realized with  four additional transversal SWAPs which induce another $\mathcal{R}_\alpha$ transformation. In the unfolded picture (i), one can make a gauge transformation to shift the blue loop to the right, in order to find correspondence with the red loop in (e). (j) The modular $\mathcal{S}$ implemented by transversal SWAPs is equivalent to two commuting cyclic permutations represented by the twist operators.  (k) One can restore the configuration in (i) by a $\pi/2$-rotation to that of the original basis in (e). The correspondence between the initial and final Wilson loops becomes clearer when viewed in this way.}	\label{fig:fold_8_layers}
\end{figure*}

\section{Transversal implementation and measurement of modular $\mathcal{S}$ on an 8-layer system}\label{append:8-layer}

Here we continue the folding procedure of Fig.~2(b) in the main text. Now we fold the system three times to form an 8-layer system, as shown in Fig.~\ref{fig:fold_8_layers}(a-d).  Importantly, note that all the periodic boundary conditions come to the left edge and can be replaced by local boundary conditions 
(gapped boundary): (16)(25)(38)(47), induced by ``gluing" the edges with local interaction (tunneling). We remark that this provides a novel way of effectively creating a torus in a planar system
with purely local interactions. 
 
One can represent the folded 8-layer system using the language of  unitary modular tensor category (UMTC) \cite{wang2008} as eight copies of UMTCs: $C \times \overline{C} \times C \times \overline{C} \times C \times \overline{C} \times C \times \overline{C}$, where $C$ describs a single copy of the original state. 
Note that the boundary conditions we specify here are such that $C$ is always glued to its parity-reversed counterpart $\overline{C}$, 
and hence can be realized with local interactions. Note that this discussion applies to both chiral 
($C \neq \overline{C}$) and non-chiral ($C=\overline{C}$) phases. 

To illustrate the scheme with clarity, we unfold the 8-layer system and label each patch with the layer number $n$, as shown in Fig.~\ref{fig:fold_8_layers}(e). 
We use $\bar{n}$ to label the layers with opposite chirality (shaded by dark blue) in the folded system, while the light-blue patches inherit the original chirality
of the unfolded system. The dashed lines delineate the lines along which the folds occur in the 8-layer system. The colored markers on the 
unfolded system indicate gluing of the opposite edges, which then correspond to local boundary conditions on the left edges of the folded system.  

In this setup, one can realize $\mathcal{R}_\alpha \mathcal{S}$ transversally through the following SWAP operations [see Fig.~\ref{fig:fold_8_layers}(f,g)]:
\be
\mathcal{R}_\alpha \mathcal{S}= \overline{\text{SWAP}}(1,2)(3,4)(5,6)(7,8).
\ee
Note that in order to illustrate the effect of the layer SWAPs on the unfolded system in Fig.~\ref{fig:fold_8_layers}, 
we show the protocol with passive transformations (change of basis), i.e., fix the location of Wilson loops but SWAP 
the patches, which is the same convention as our definition of modular transformations in Fig.~1 (a-c) 
in the main text. Meanwhile, we illustrate active transformations on the folded systems, i.e., change the location 
(layer labels) of the Wilson loops. The $\mathcal{R}_\alpha \mathcal{S}$ transformation corresponds to a mirror 
reflection along the diagonal line: $M_{\hat{x} - \hat{y}}$ as shown in Fig.~\ref{fig:fold_8_layers}(g). One can see that 
in both the folded and unfolded pictures, the red loop has the path $2\rightarrow 1 \rightarrow 8 \rightarrow 5 \rightarrow 2$, 
which was the original path of the blue loop in (e) with the opposite direction, indicating the transformation $\mathcal{R}_\alpha \mathcal{S} : (\alpha, \beta) \rightarrow (-\beta, -\alpha)$. 

One can also apply an additional reflection transversally by $\mathcal{R}_\alpha= \overline{\text{SWAP}}(1,8)(2,5)(3,6)(4,7)$ 
after the $\mathcal{R}_\alpha \mathcal{S}$ transformation, in order to cancel out the reflection, as shown in 
Fig.~\ref{fig:fold_8_layers}(i,k). This leads to the realization of modular $\mathcal{S}$ transversally. In the 
unfolded system in Fig.~\ref{fig:fold_8_layers}(i), $\mathcal{R}_\alpha$ corresponds to a mirror reflection along the vertical line 
in the new basis: $M_{\hat{y}'}$ (equivalent to $M_{\hat{x}}$ in the original basis).  We can see that the red loop has 
the path $5 \rightarrow 8 \rightarrow 1 \rightarrow 2 \rightarrow 5$, which has exactly the same path as the original blue loop $W^\alpha_a$ in (e), i.e., $2 \rightarrow 1 \rightarrow 8 \rightarrow 5 \rightarrow 2 $. 
This indicates the red loop in  (i) is $W^{\alpha}_b$. 
One can see this correspondence more clearly when restoring  the orientation of the manifold by a 
$\frac{\pi}{2}$-rotation [Fig.~\ref{fig:fold_8_layers}(k)] to the configuration of the original basis in (e). 
On the other hand, the blue loop in (i) has the path $8 \rightarrow 5 \rightarrow 4 \rightarrow 3 \rightarrow 8$, 
which is gauge-equivalent to the path $1 \rightarrow 2 \rightarrow 7 \rightarrow 6 \rightarrow 1$. This is 
exactly the path of the original red loop $W^\beta_b$ in (e) with the opposite direction, as can also be seen in the restored configuration (k). 
This indicates that the blue loop in (i) is $W^{-\beta}_a$. Therefore the whole transformation from (d,e) to (h,i) 
achieves $\mathcal{S}:(\alpha, \beta) \rightarrow (-\beta, \alpha)$.

The transversal modular $\mathcal{S}$ can also be abbreviated by two commuting cyclic twist operators as
\be
\mathcal{S}=\overline{V}(1735)\overline{V}(2648).
\ee
To be concrete, $\overline{V}(1735)$ represents the cyclic permutation $(1\rightarrow 7 \rightarrow 3 \rightarrow 5 \rightarrow 1)$. The equivalence to two commuting twist operations is illustrated in Fig.~\ref{fig:fold_8_layers}(j).  Note that a single twist operator only couples layers with the same chirality, as required. 

We note that the folding procedure and transversal operation discussed here have some resemblance 
with the interpretation of a color code as folded surface code \cite{Kubica:2015br} and the existence 
of transversal Clifford gate set in the triangular color code \cite{Bombin:2006hw}. However here we are considering
a torus in terms of a square with periodic boundary conditions (as opposed to a square with gapped boundaries), 
and our construction applies to arbitrary Abelian and non-Abelian topological states. 

\begin{figure} 
  \includegraphics[width=1\columnwidth]{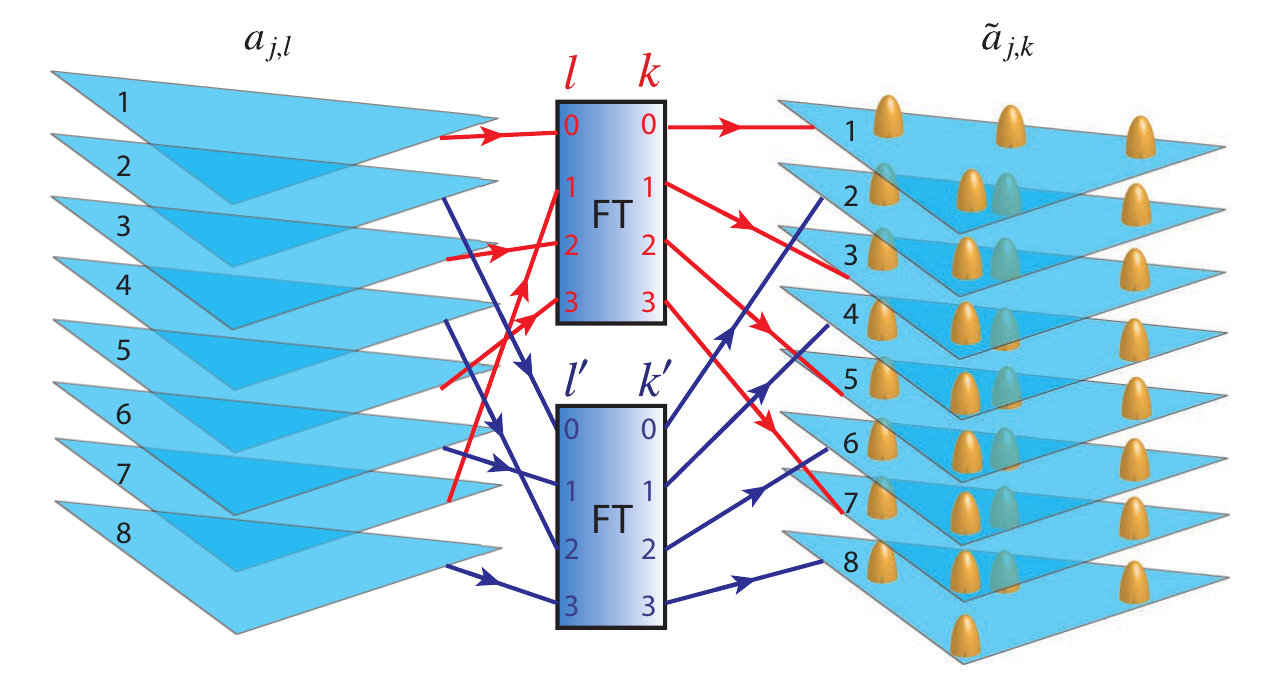}
  \caption{ Measuring $\mathcal{S}$ as a general permutation on a 8-layer system using two quantum Fourier 
transform gadgets and readout in the Fourier basis.  The FT gadget can be implemented by a sequence of beamsplitter operations.}
  \label{fig:meausre_twist}
\end{figure}

The measurement of the two twist operators, $\overline{V}(1735) \overline{V}(2648)$, proceeds according to the general discussion
presented in Sec.~\ref{sec:measure_permutation} (see Fig.~\ref{fig:meausre_twist} for an illustration). We apply Fourier transforms (FT) to 
layer 1,7,3,5 and 2,4,6,8 (sorted according to the order when entering the input $l$ and $l'$ of the FT circuit gadget) 
respectively as illustrated in Fig.~\ref{fig:meausre_twist}(b).  The FT gadget performs
the linear map in Eq.~(11), and then one measures the twist operators using the relation in Eq.~(12). 

\begin{figure} 
  \includegraphics[width=1\columnwidth]{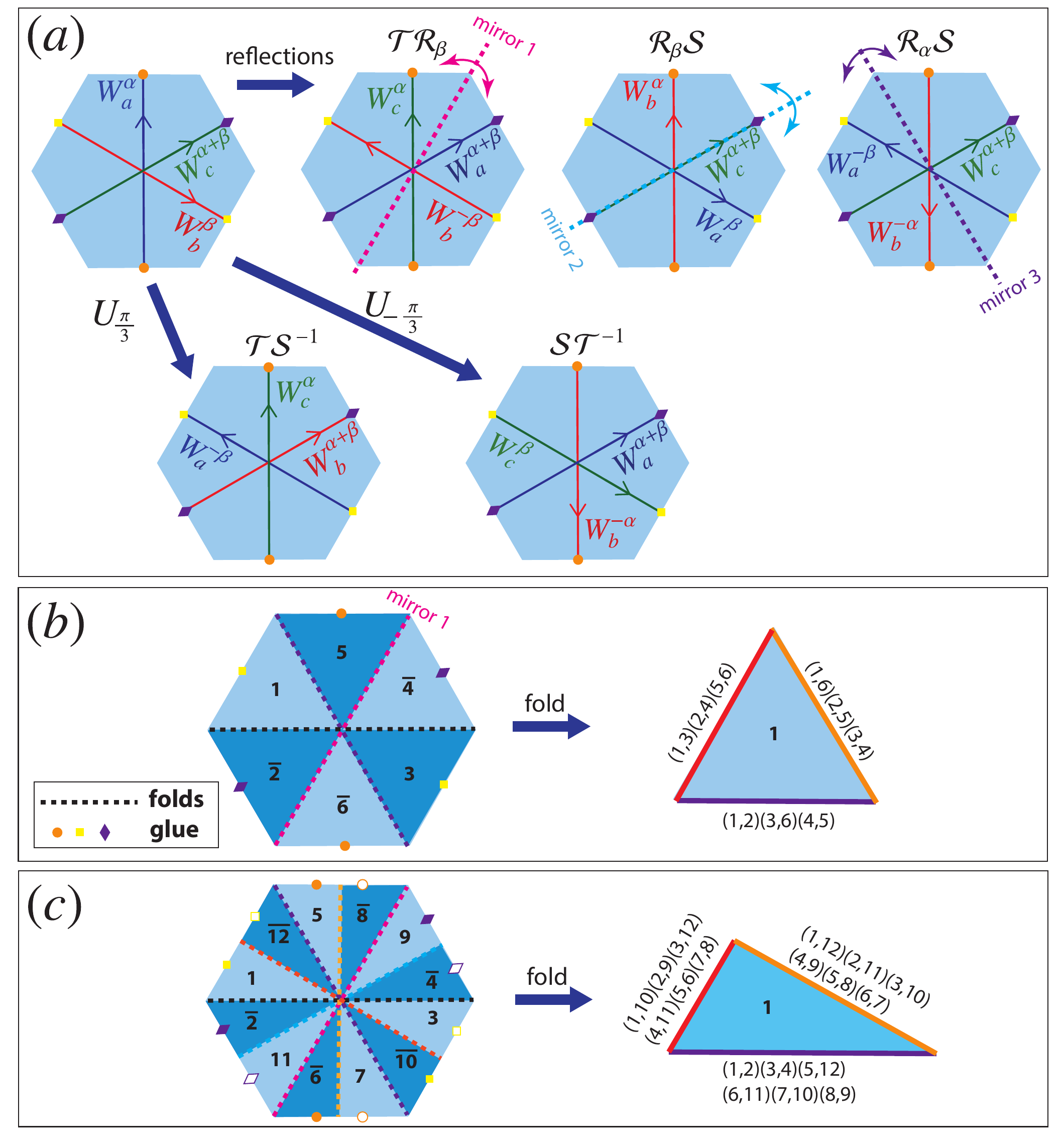}
  \caption{(a)  A hexagon surface with opposite edge being identified is equivalent to a torus.  The reflections along different axis implements transformations $\mathcal{T}\mathcal{R}_\beta$, $\mathcal{R}_\beta \mathcal{S}$, $\mathcal{R}_\alpha \mathcal{S}$.  Rotation for $\pm \frac{\pi}{3}$ implements $\mathcal{T}\mathcal{S}^{-1}$ and $\mathcal{S}\mathcal{T}^{-1}$ respectively. (b) Folding the hexagon into 6 layers lead to a triangle code with gapped boundaries on the edge. The reflections shown above can be implemented transversally in this system.  (c) All the reflections and rotations shown in (a) can be implemented transversally on the same code when folding the hexagon into 12 layers.}
  \label{fig:hexagon_lattice}
\end{figure}

\begin{figure*}
  \includegraphics[width=2\columnwidth]{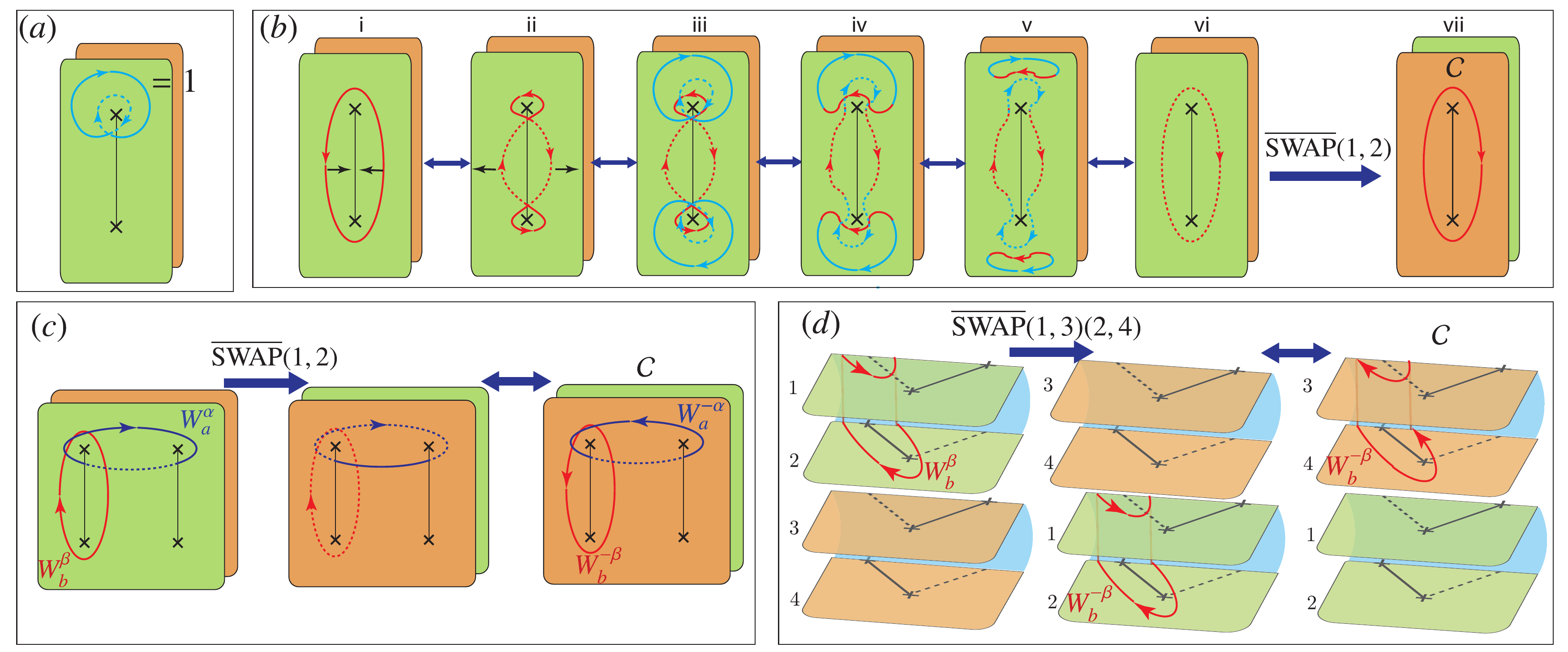}
  \caption{Wilson loop relations and charge conjugation.  Dashed lines in the figures using top view represent anyons travelling to the bottom layer. 
(a)  The double loop around the genon is considered to be contractible and equivalent to identity. 
(b) Derivation of the Wilson-loop relation that a loop is equivalent to the loop in the opposite layer with opposite direction.  
The relation in panel (a) has been used for the derivation. Transversal SWAP between the two layers flips the direction of the 
loop and hence is equivalent to charge conjugation operator.  (c)  On a bilayer system with two pairs of genons (equivalent to an effective torus), 
interlayer transversal SWAP implements a charge conjugation operation that flips the direction of the loop in both the $\alpha$- and $\beta$-cycle. 
(d)  When folding the bilayer system in (c) into a 4-layer system, interlayer SWAP between layer 1 and 3, and layer 2 and 4  implements the charge conjugation.}
\label{fig:loop_relations}
\end{figure*}

\begin{figure*}[hbt]
  \includegraphics[width=1.6\columnwidth]{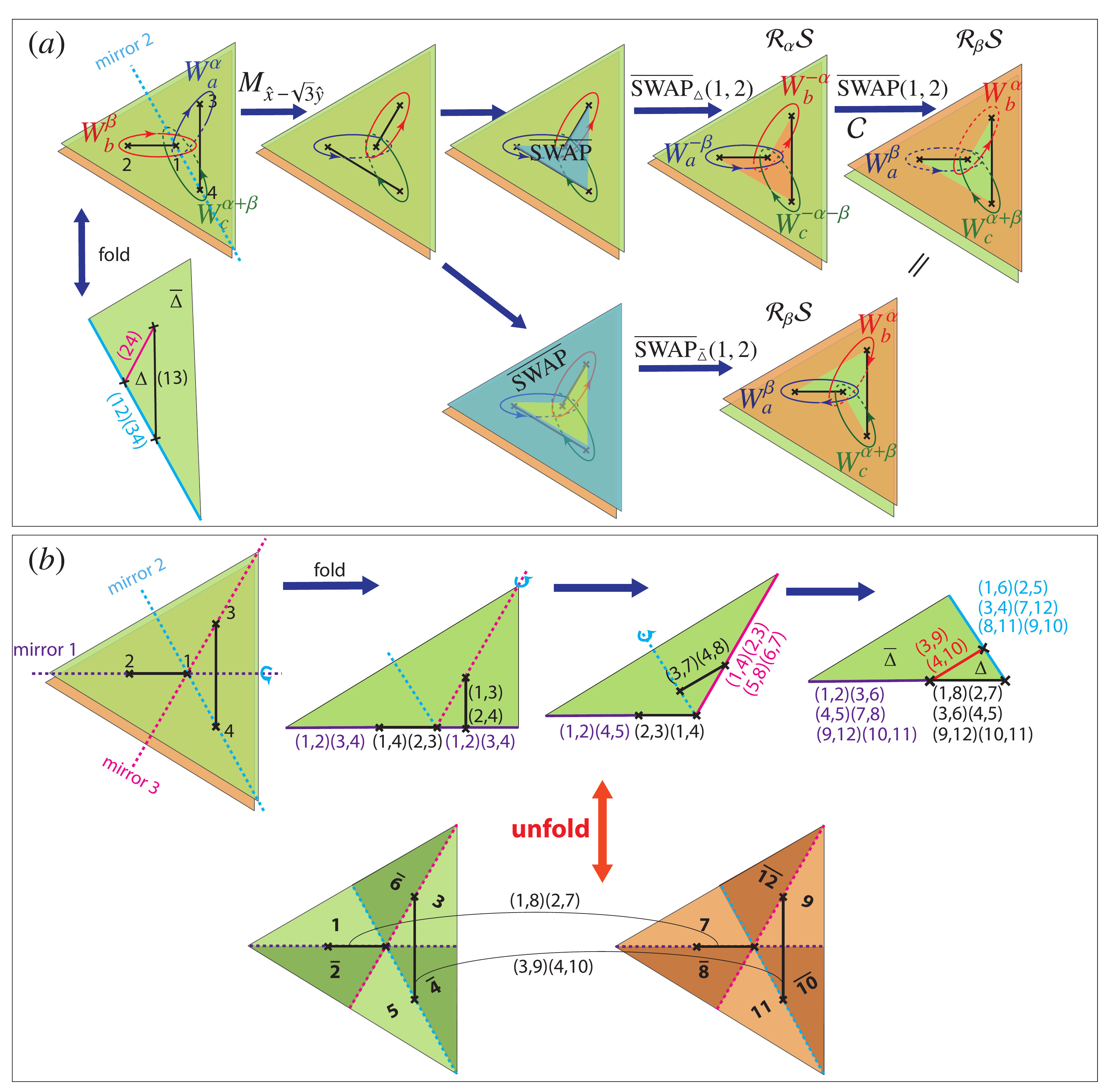}
  \caption{(a) Implementing $\mathcal{R}_\alpha \mathcal{S}$ and $\mathcal{R}_\beta \mathcal{S} $ with reflection along the mirror axis 
connecting defect 1 and 4.   All the protocols can be applied transversally when folding the system into 4 layers along the mirror. 
(b) By folding the systems along the three types of mirrors into 12 layers, one can achieve both 
$\mathcal{R}_\beta \mathcal{S}$ and $ \mathcal{T}\mathcal{R}_\beta$ transversally without moving the genons. 
The gapped boundaries and twists are labeled on the folded system.  The correspondence to the 
unfolded system is shown on the bottom of the panel and label the layers on each patch with the same 
rules as in Fig.~\ref{fig:fold_8_layers}.  The bar on the number along with the dark shade indicates layers 
with reversed chirality (orientation) in the folded system. The branch cuts connecting the green and orange patches are 
indicated by the lines and corresponding labels. }
\label{fig:triangle_12_layers}
\end{figure*}

\begin{figure}[hbt]
  \includegraphics[width=1\columnwidth]{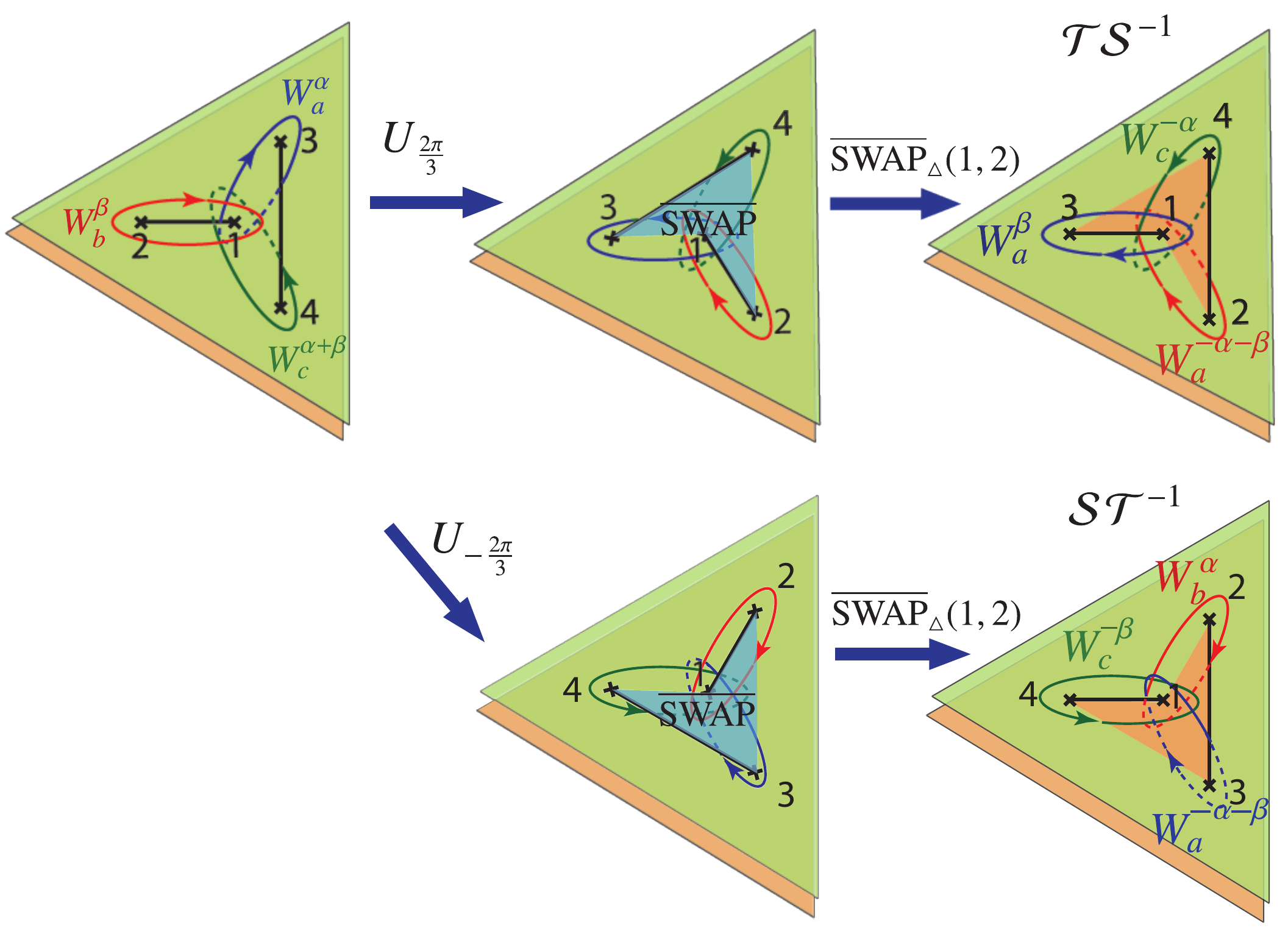}
  \caption{$\pm 2\pi/3$ rotations combined with SWAPs inside the central patches  
on a $C_{3v}$ symmetric geometry of genons implements modular $\mathcal{T}\mathcal{S}^{-1}$ and 
$\mathcal{S}\mathcal{T}^{-1}$ transformations respectively.}
\label{fig:triangle_rotation}
\end{figure}

\section{Transversal implementation of modular transformations on a torus with hexagonal symmetry}\label{append:hexagon}

In the previous section , we have shown that a torus with $C_4$-rotational symmetry  can be used to  implement modular $\mathcal{S}$, $\mathcal{R}_\alpha$, $\mathcal{R}_\beta$ and their combinations transversally by folding the manifold into 8 layers.    In this section, we show that a torus with $C_6$ symmetry can be used to implement modular transformations related to $\mathcal{T}$.

As shown in the upper-left panel in Fig.~\ref{fig:hexagon_lattice}(a), a hexagon with the three pairs of opposite edges being identified is equivalent to a torus.   There are two independent loops $\alpha$ (blue) and $\beta$ (red) on the manifold, and the third one $\alpha+\beta$ (green) is their superposition.   As one can see, a reflection along `mirror 1' implements the map $\mathcal{T}\mathcal{R}_\beta:(\alpha, \beta) \rightarrow (-\beta, \alpha)$.     Similarly, the reflections along `mirror 2' and `mirror 3' implements $\mathcal{R}_\beta \mathcal{S}$ and $\mathcal{R}_\alpha \mathcal{S}$.  The combinations of reflections can also implement rotations, for example, $U_{\pi/3}$ and $U_{-\pi/3}$ as shown in the lower panel of Fig.~\ref{fig:hexagon_lattice}(a), which gives rise to $\mathcal{T}\mathcal{S}^{-1}$ and $\mathcal{S}\mathcal{T}^{-1}$.

Each individual reflection can be implemented transversally if folding the manifold 3 times and into 6 layers, as illustrated in Fig.~\ref{fig:hexagon_lattice}(b), where the $\mathcal{T}\mathcal{R}_\beta$ can be implemented transversally.  The corresponding gapped boundaries created by folding and gluing the edges is shown explicitly in the right panel of (b).  Furthermore, one can implement all the modular transformations shown in (a) transversally on the same code by folding the manifold into 12 layers as shown in (c).

\section{Transversal topological charge conjugation and Wilson loop relations in the presence of genons}\label{append:loop-relation}

Here we explain some relations involving Wilson loops in the presence of genons, and use them to implement 
the topological charge conjugation operation $\mathcal{C}$. 

The first relation is shown in Fig.~\ref{fig:loop_relations}(a): the double loop encircling a single genon twice is 
contractible (in the mapping to the effective high genus surface, it represents a contractible loop). Thus the associated 
Wilson loop operator for an anyon encircling such a double loop can be taken to act as the identity on the 
ground state subspace. More precisely, one can consider fusing an arbitrary anyon with the genon to obtain a new type of genon, in which case the eigenvalue
of this Wilson loop will change \cite{barkeshli2014SDG}. However there always exists a reference genon for which the double Wilson loop is
the identity, and this is the one we work with. In the case of stabilizer codes such as surface code, the double loop is just a single stabilizer 
operator around the defect, which can be set to $1$ by measurement. 

Using this first relation, we can derive the second relation: in the bilayer system with genons,
a non-contractible cycle is equivalent to itself under the combined action of layer exchange 
and reversal of the direction (flipping the arrow). We derive this relation step by step as shown in Fig.~\ref{fig:loop_relations}(b).  
In (i), we stretch both the left and right parts of the loops in layer 1 across the branch cut such that the anyon lines 
travel to the opposite layer (shown by dashed lines) and cross each other as shown in (ii). Now in (iii) we multiply the 
double loops (equal to $1$) around both the top and bottom genons.  In (iv) we show that the double loops can be recombined 
with the original loops in the same layers.  We locally deform and detach the three loops in (v).  One can see both the top 
and bottom loops are contractible and equal to $1$. Finally, in (vi), we end up with a single loop in the opposite layer and 
with an opposite direction as the loop in (i). Through the previous steps, we have derived that the Wilson loops in (i) and 
(vi) are identical. Now in (vii), we SWAP the loop back to layer 1.  When comparing the original loop in (i), we see that 
the direction of the loop is flipped. 

Now let us consider two pair of genons, forming an effective torus as shown in (c).  In this case, we can see that 
the transversal SWAP between layer 1 and 2 flips the directions of both loops, i.e., 
$\overline{\text{SWAP}}(1,2) : (\alpha, \beta) \rightarrow (-\alpha, -\beta)$.  This is exactly the 
definition of the charge conjugation operator $\mathcal{C}$ in Appendix \ref{append:MCG}.   
Therefore, in the bilayer system with genons, $\mathcal{C}=\overline{\text{SWAP}}(1,2)$. 

Finally, when we fold the bilayer system into the 4-layer system in (d), the charge conjugation operation becomes 
$\mathcal{C}$$=$$\overline{\text{SWAP}}(1,3)(2,4)$, i.e., transversal SWAP between layer 1 and 3, and layer 2 and 4.  
Note that this SWAP preserves the gapped boundary (fold), therefore mapping the system back to the original Hilbert space~$\mathcal{H}$.  

\section{Protocols on triangular defect code implementing both $\mathcal{R}\mathcal{S}$ and $\mathcal{T}\mathcal{R}$ transversally}\label{append:12-layer}

In the main text, we have shown how to perform $\mathcal{T} \mathcal{R}_\beta $ transformation 
transversally in the triangular configuration of genons in Fig.~3(c).  We also
showed how to perform $\mathcal{R}_\beta \mathcal{S}$ by switching to the genon configuration with $C_{4v}$ symmetry. 
Therefore implementing both transformations in a 4-layer system requires moving defects. 

Here, we show explicitly that $\mathcal{R}_\beta \mathcal{S}$, $\mathcal{T}\mathcal{R}_\beta$, and $\mathcal{C}$ 
can all be performed transversally with layer permutations, without moving the defects at all, by folding the system 
into a 12-layer system. 

The scheme is shown in Fig.~\ref{fig:triangle_12_layers}. In panel (a), we first apply 
a reflection along the line connecting defect 1 and 4, i.e.,~$M_{\hat{x}-\sqrt{3}\hat{y}}$, 
followed by a SWAP operation in the blue shaded regions, $\overline{\text{SWAP}_\triangle}(1,2)$ to return the branch cuts
to their original configurations. This achieves an $\mathcal{R}_\alpha \mathcal{S}$ transformation, which acts on the loop as
$\mathcal{R}_\alpha \mathcal{S}:(\alpha, \beta)\rightarrow (-\beta, -\alpha)$. In order to also realize $\mathcal{R}_\beta \mathcal{S}$, 
one can apply an additional global SWAP, i.e., $\overline{\text{SWAP}}(1,2)$, equivalent to the topological 
charge conjugation operation $\mathcal{C}$, as was described above in Appendix \ref{append:loop-relation}. This in turn achieves the mapping 
$\mathcal{R}_\beta \mathcal{S}:(\alpha, \beta)\rightarrow (\beta, \alpha)$, which interchanges the red and blue loop 
in the initial (leftmost) configuration. One can conclude the following relation: $\mathcal{R}_\beta \mathcal{S}= \mathcal{C}\mathcal{R}_\alpha \mathcal{S}$. To see the correspondence between the initial configuration more clearly, 
we note that reversing the arrows in the loops is equivalent to switching the layers of the loop (as shown in the 
rightmost configuration in the second row and explained in Appendix \ref{append:loop-relation}).  
One can also directly achieve $\mathcal{R}_\beta \mathcal{S}$ by performing the initial inter-layer SWAP between 
the blue shaded regions outside the central patch, i.e., $\overline{\text{SWAP}}_{\bar{\triangle}}(1,2)$, as shown in the second row of panel (a).   

All of the above operations can be performed transversally by layer SWAPSs in a 4-layer system when 
folding along the mirror axis connecting defect 1 and 4, as shown in the lower-left part of panel (a). 
We can hence express $\mathcal{R}_\alpha \mathcal{S}$ and $\mathcal{R}_\beta \mathcal{S}$ as
\begin{align}
\non \mathcal{R}_\alpha \mathcal{S} =&\overline{\text{SWAP}}_\Delta (1,4)(2,3)\overline{\text{SWAP}}_{\bar{\Delta}} (1,2)(3,4)  \\
 \mathcal{R}_\beta \mathcal{S} =&   \overline{\text{SWAP}}_\Delta (1,2)(3,4)\overline{\text{SWAP}}_{\bar{\Delta}} (1,4)(2,3). 
\end{align}
Here, as in the discussion of the $C_{4v}$ geometry in the main text, $\Delta$ refers to the
triangular region enclosed by the branch cuts after folding, and $\overline{\Delta}$ refers to its complement. 

In the main text, we have already shown that folding along the horizontal mirror axis connecting defect 1 and 2 (into 4 layers) 
enables applying $\mathcal{T}\mathcal{R}_\beta $ transversally.  Here, we show that by folding the whole system along 
the three types of mirror axes (with an additional one connecting defect 1 and 3) shown in Fig.~\ref{fig:triangle_12_layers}(b), 
one can get a 12-layer system  with three types of gapped boundaries and one branch cut in the bulk.  To illustrate the 
correspondence with the original bilayer system, we unfold system at the bottom of panel (b), with the 
layer labels on each patch.   The bar on top of the number and shade of the patch indicates that after folding, that region has
reversed orientation (chirality) relative to the original system. The mirrors (dashed lines) on the unfolded system
separating layers with opposite chirality correspond to gapped boundaries in the folded system.  The horizontal twist 
in the unfolded system overlaps with one of the mirror axes and hence modifies the gapped boundary in the lower-right 
part of the folded system.  The vertical branch cut in the unfolded system is located in the bulk of the folded 12-layer system 
(represented by the red line).     

In such a folded system, one can perform $ \mathcal{T}\mathcal{R}_\beta$, $\mathcal{R}_\beta \mathcal{S}$, and $\mathcal{C}$ 
transversally without moving genons:
\begin{align}
\non  \mathcal{T} \mathcal{R}_\beta=& \overline{\text{SWAP}} (1,2)(3,4)(5,6)(7,8)(9,10)(11,12) \\
\non \mathcal{R}_\beta \mathcal{S} =& \overline{\text{SWAP}}_\Delta(1,6)(2,3)(4,5)(7,12)(8,9)(10,11) \\
\non  & \overline{\text{SWAP}}_{\bar{\Delta}}(1,12)(2, 9)(3, 8)(4, 11)(5, 10)(6, 7).  \\
\mathcal{C} =& \overline{\text{SWAP}}_{\bar{\Delta}}(1,7)(2,8)(3,9)(4,10)(5,11)(6,12).
\end{align}
Here, we again denote the triangular patch with three genons at its vertices at the lower right corner of the folded 12-layer system 
by $\Delta$, and its complement by $\bar{\Delta}$.  Note that $\mathcal{C} = \mathcal{R}_\alpha \mathcal{R}_\beta = \mathcal{R}_\beta \mathcal{R}_\alpha$.   Therefore, one can easily get $\mathcal{T}\mathcal{R}_\alpha = \mathcal{T}\mathcal{R}_\alpha \mathcal{C} $ and $\mathcal{R}_\alpha \mathcal{S}= \mathcal{C} \mathcal{R}_\beta \mathcal{S}$ by transversally applying an additional $\mathcal{C}$ as mentioned above. 

Note that one can also apply combinations of such transformations, which allows transversal implementation of 
$\mathcal{T}\mathcal{S}=(\mathcal{T}\mathcal{R}_{\beta})(\mathcal{R}_{\beta} \mathcal{S}) $, $(\mathcal{T}\mathcal{S})^{-1}=(\mathcal{R}_{\beta} \mathcal{S})(\mathcal{T}\mathcal{R}_{\beta}) $ and 
$ \mathcal{T} \mathcal{S}^{-1} =( \mathcal{T}\mathcal{R}_\beta)( \mathcal{R}_\alpha \mathcal{S})$, etc. 
Note that $\mathcal{T}\mathcal{S}$, $(\mathcal{T}\mathcal{S})^{-1}$ and $ \mathcal{T} \mathcal{S}^{-1} $ are all elements of the orientation-preserving mapping class group,
and hence they preserve the ground state subspace of chiral topological phases. 

We note that a sequence of reflections about different mirror axes implements a $\pm 2\pi/3$ rotation of the unfolded geometry.
Therefore, the $\pm 2\pi/3$ rotation combined with SWAPs in the unfolded system implements $\mathcal{T}\mathcal{S}^{-1}$  
and  $\mathcal{S}\mathcal{T}^{-1}$ transformations respectively, as shown in Fig.~\ref{fig:triangle_rotation}.   Applying global SWAPs between two layers on top the above transformations implement an additional charge conjugation $\mathcal{C}$, giving rise to $\mathcal{T}\mathcal{S}$ and $(\mathcal{T}\mathcal{S})^{-1}$ respectively.  
This again illustrates the fact that spatial symmetry transformations can act as protected logical operations.

\section{Examples of fault-tolerant gates}\label{append:fault-tolerant}

Table II in the main text listed several examples of modular matrices and the mapping class group of high genus surfaces
for different topological states. Here we briefly elaborate on some of the simplest examples. 
 
 (1) $\nu = 1/k$ \it{ Laughlin fractional quantum Hall state}: \rm
We begin by discussing topological states whose topological order coincides with that of the $\nu = 1/k$ Laughlin fractional quantum Hall (FQH) states.  

In the simplest case of $\nu=1/2$ bosonic FQH state, the ground-state degeneracy on a torus is 2, encoding  one logical qubit labeled 
as $\ket{n_s}_\beta$. Here, $n_s=0,1$ represents the semion number, which measured by applying a Wilson loop 
along the $\alpha-$cycle.  The corresponding logical Pauli operators are $\overline{Z}=W^{\alpha}_s$ and $\overline{X}=W^{\beta}_s$.
The modular transformations 
 \be
\non \mathcal{S} W^{\beta}_{s} \mathcal{S}^\dag = W^{\alpha}_s, \quad \mathcal{S} W^{\alpha}_{s} \mathcal{S}^\dag = W^{-\beta}_s=W^{\beta}_{\mathcal{C}(s)}=W^{\beta}_s
 \ee
  and 
  \be
\non \mathcal{S} W^{\beta}_{s}W^{\alpha}_{s} \mathcal{S}^\dag=\mathcal{S} W^{\beta}_{s} \mathcal{S}^\dag\mathcal{S}  W^{\alpha}_{s} \mathcal{S}^\dag=W^{\alpha}_s W^{-\beta}_{s}=W^{\alpha}_s W^{\beta}_{s}=-W^{\beta}_s W^{\alpha}_{s}
 \ee
 are equivalent to 
 \be
 \overline{H}  \ \overline{X} \ \overline{H} = \overline{Z},\quad \overline{H}  \ \overline{Z} \ \overline{H} = \overline{X}\quad  \text{and} \quad \overline{H}  \ \overline{Y} \ \overline{H} = -\overline{Y}.
 \ee 
That is, the action of $\mathcal{S}$ is equivalent to that of a logical Hadamard gate $\overline{H}=\frac{1}{\sqrt{2}}\left(^{1\ \ 1}_{1 \ \ -1} \right)$.  
Note that we have used the property that semion is self-conjugate, i.e., $\mathcal{C}(s) = s$, where $\mathcal{C}$ is the 
topological charge conjugation operator.  The modular $\mathcal{T}$ has the matrix representation in the computational 
basis as $\mathcal{T}$$=$$ \left( ^{1 \ \ 0}_{0 \ \  i} \right)$$=$$\overline{P}$, equivalent to the logical Clifford phase gate $\overline{P}$.  

One can generalize the above analysis to the $\nu=1/k$ Laughlin FQH state, where the ground-state degeneracy on a torus 
is $k$, corresponding to a $k$-level logical qudit, labeled as $\ket{a}_\beta$, where $a=0, 1, 2, ..., $. In this case, the 
Wilson loops of inserting a single anyon along the two cycles of the torus play the role of qudit shift operators, i.e.,  
$\overline{Z}=W^{\alpha}$ and  $\overline{X}=W^{\beta}$, with $\overline{X} \overline{Z} = e^{2\pi i/k} \overline{Z} \overline{X}$. 
The shift operator acts on the basis state as $\overline{X} \ket{a}_\beta = \ket{a+1\ \text{mod}\ k}_\beta$ and $\overline{X}^\dag \ket{a}_\beta = \ket{a-1\ \text{mod}\ k}_\beta$.  
In this case, the modular $\mathcal{S}$ has the matrix representation in the computational basis as 
$\mathcal{S}_{ab}=\frac{1}{\sqrt{k}}e^{i2\pi a b/k}$, which is a discrete Fourier transform of the basis states.  
The modular $\mathcal{T}$ has the representation  $\mathcal{T}_{ab}=\delta_{ab}e^{i2\pi a(a+k)/2}$, which is a generalized phase gate on the logical qudit. 

In the case of $\nu=1/2$ bosonic FQH, the whole $\text{MCG}_{\Sigma}$ of a genus $g$ surface
$\Sigma$ generates the Clifford group for the logical qubits. In the general $\nu=1/k$ case, 
the $\text{MCG}_{\Sigma}$ generates the generalized Clifford group for the logical qudits.

(2) \it{Double semion state}: \rm The topological order of the double semion coincides with that of
two decoupled layers of $\nu$$=$$\pm 1/2$ bosonic FQH states, where the layers have opposite chirality. 
The complete set of Wilson operators are the semion ($s$) and anti-semion ($s'$) loops, 
with the pair of anti-commutation relations $\{W^{\alpha}_s, W^{\beta}_s\}=\{W^{\alpha}_{s'}, W^{\beta}_{s'}\}=0$.
Each particle is self-conjugate, as $s \times s = s' \times s' = \mathbb{I}$, where $\mathbb{I}$ denotes the trivial (vacuum) sector. 
The Wilson loops associated with different semions, $s$ and $s'$, commute with each other. 
The ground states can be labeled by the number of  loops inserted along the $\beta$-direction, i.e., 
$\ket{n_s, n_{s'}}_{\beta} =(W^{\beta}_s )^{n_s} ( W^{\beta}_{s'})^{n_{s'}}\ket{0_s, 0_{s'}}_{\beta} $, with $n_s, n_{s'} = 0,1$.  

We can represent the topological ground state subspace (code space)
as a tensor product of the semion and anti-semion parts, i.e.,
$\mathcal{H} =\mathcal{H}_{\frac{1}{2}} \otimes \mathcal{H}_{-\frac{1}{2}}$. We can choose to store the 
logical qubit information only in the semion subspace $\mathcal{H}_{\frac{1}{2}}$ by tracing out (ignoring) the anti-semion part 
(considering the situation where there is no entanglement between semion and anti-semion), i.e., $\mathcal{H}_{\frac{1}{2}}=\text{Tr}_{-\frac{1}{2}}\mathcal{H}$.
Thus we can label the logical qubit by the semion number $\ket{n_s}_{\beta}$.  The corresponding logical operators 
are $\overline{Z}=W^{\alpha}_s$ and $\overline{X}=W^{\beta}_s$, which is the same as the $\nu=1/2$ FQH states.  Therefore, 
the previous discussion applies here as well. The modular $\mathcal{S}$ and $\mathcal{T}$ applies logical Hadamard 
$\overline{H}=\frac{1}{\sqrt{2}}\left(^{1\ \ 1}_{1 \ \ -1} \right)$ and phase $\overline{P}=\left(^{1\ \ 0}_{0 \ \ i} \right)$ respectively, and $\text{MCG}_{\Sigma}$ generates the Clifford group.

Now we also consider $\mathcal{R}_{\alpha} \mathcal{S}$ and $ \mathcal{T} \mathcal{R}_{\alpha}$ transformation, 
which can be implemented transversally with only 4 layers. For simplicity, we could instead first consider the 
transversal logical gate corresponding to $\mathcal{R}_\alpha$, i.e.,
 \be
 \mathcal{R}_{\alpha}W^{\alpha}_s \mathcal{R}_{\alpha} =W^{-\alpha}_{s'}=W^{\alpha}_{\mathcal{C}(s')}=W^{\alpha}_{s'} \quad \text{and} \quad \mathcal{R}_{\alpha} W^{\beta}_s \mathcal{R}_{\alpha} =W^{\beta}_{s'}. 
 \ee
 One can see that the reflection $\mathcal{R}_\alpha$ exchanges semion and anti-semion.  This can again be intuitively 
understood from the picture of two copies of  FQH states with opposite chirality, since the reflection changes the chirality 
of both copies and hence turns semion into anti-semion, and vice versa.   Therefore, if the logical quantum information is stored in the semion sector
$\mathcal{H}_{\frac{1}{2}}$, the reflection $\mathcal{R}_\alpha$ takes the semion to anti-semion and hence takes the logical 
qubit out of the logical subspace.  Therefore,  $\mathcal{R}_\alpha \mathcal{S}$ or $ \mathcal{T}\mathcal{R}_\alpha$ alone 
is not considered as a logical gate in this encoding scheme.   Only when another $\mathcal{R}_\alpha$ is applied and 
the combination leads to $\mathcal{S}=\mathcal{R}_\alpha(\mathcal{R}_\alpha\mathcal{S})$, 
$\mathcal{T}=(\mathcal{T}\mathcal{R}_\alpha)\mathcal{R}_\alpha$,  or 
$\mathcal{T}\mathcal{S}=(\mathcal{T}\mathcal{R}_{\alpha})(\mathcal{R}_{\alpha} \mathcal{S}) $ etc., 
do we get a logical gate.  This consideration holds for the other doubled states as well, such as the 
$\text{Ising} \times \overline{\text{Ising}}$ and $\text{Fib.} \times \overline{\text{Fib.}}$ states, which 
can both be considered as two copies of topological states with opposite chirality. 
 
 (3) \it{ $\mathbb{Z}_2$ Toric code ($\mathbb{Z}_2$ spin liquid)}: \rm 
The complete set of Wilson operators are the spinon ($e$) and vison ($m$) 
loops on both cycles, satisfying the anti-commutation relation $\{W^{\alpha}_e, W^{\beta}_m\} =\{W^{\beta}_e, W^{\alpha}_m\}=0$. 
We can choose the logical qubit basis as $\ket{n_e n_m}_{\beta}$ and the corresponding logical operators become 
$\overline{Z}_{1,2}=W^{\alpha}_{e,m}$ and $\overline{X}_{1,2}=W^{\beta}_{m,e}$.

Modular $\mathcal{S}$ transforms the logical operators as $\mathcal{S} (W^{\alpha}_{e})^n (W^{\alpha}_{m})^l  \mathcal{S}^\dag $$=$$ (W^{\beta}_{e})^n (W^{\beta}_{m})^l $ (with $n,l=0,1$).
This is equivalent to $\overline{\text{SWAP}}_{12} \overline{H}_1 \overline{H}_2 \left(\overline{Z}_1\right)^n \left(\overline{Z}_2\right)^l \overline{H}_1 \overline{H}_2 \overline{\text{SWAP}}_{12} $$=$$ \left(\overline{X}_2\right)^n \left(\overline{X}_1\right)^l$, i.e., logical Hadamard gates on both qubits with an additional logical SWAP.  The modular $\mathcal{T}$ in this basis has the representation as $\mathcal{T}=\text{diag}(1, 1,  1, -1)$, corresponding to a $\overline{CZ}$ gate.   
  
Note that, for Kitaev's $\mathbb{Z}_2$ toric code model, one can choose a particular reflection axis such that the anyons transform 
trivially under reflection: $\mathcal{R}_\alpha(e)=e$ and $\mathcal{R}_\alpha(m)=m$, and similarly for $\mathcal{R}_\beta$. 
In this case, the reflection makes the following transformation 
\be
 \mathcal{R}_{\alpha}W^{\alpha}_e \mathcal{R}_{\alpha} =W^{-\alpha}_{\mathcal{R}(e)}=W^{-\alpha}_{e}=W^{\alpha}_{\mathcal{C}(e)}=W^{\alpha}_{e}.
\ee
Similar relations hold for $W^{\alpha}_m$.  Therefore, $\mathcal{R}_\alpha$ acts trivially on the states. Thus for this topological 
order, $\mathcal{R}_\alpha\mathcal{S}$ is equivalent to $\mathcal{S}$, and $\mathcal{T}\mathcal{R}_\alpha$ is equivalent to $\mathcal{T}$. 

 \vspace{0.1in}

\section{Extracting off-diagonal  matrix elements and state preparation for measurement protocols}\label{sec:off-diagonal}

\nin We have shown above how one can measure expectation values of modular transformations and the diagonal matrix elements of them. In order to fully diagnose topological order, 
and to fully determine the fractional statistics of the quasiparticles, we need to measure all of the matrix elements of $\mathcal{S}$ and $\mathcal{T}$. 
This is a difficult problem in general; below we discuss some potential methods for state preparation, in order to reliably access desired matrix
elements.

In order to also measure the off-diagonal elements of $\mathcal{S}$, e.g., ${_{\beta}}\boket{a}{\mathcal{S}}{b}_{\beta}$, 
one should prepare the many-body state in a superposition, i.e.~$\ket{\psi}=\frac{1}{\sqrt{2}}[\ket{a}_{\beta} + \ket{b}_{\beta}]$,
and then measure the expectation value as  
\begin{align}\label{eq:off-diagonal}
\non \boket{\psi}{\mathcal{S}}{\psi}  =& \frac{1}{2} \left[{_{\beta}}\boket{a}{\mathcal{S}}{a}_{\beta} + {_{\beta}}\boket{b}{\mathcal{S}}{b}_{\beta} \right] \\
 &+ \frac{1}{2}{_{\beta}}\boket{a}{(\mathcal{S}+\mathcal{S}^\dag)}{b}_{\beta},
\end{align}
In general, $\mathcal{S}^\dagger = \mathcal{C} \mathcal{S}$, where $\mathcal{C} = \mathcal{S}^2$ is the charge conjugation operator, whose
action on loops is: $\mathcal{C}:(\alpha, \beta) \rightarrow (-\alpha, -\beta)$. Thus $\mathcal{S}=\mathcal{S}^\dag$ when $\mathcal{C}$ can 
be treated as the identity operation, which happens in states where each anyon is its own anti-particle: $\mathcal{C}:a \rightarrow \bar{a}=a$.
This is the case, for example, in the $\mathbb{Z}_2$ spin liquid, double semion, Ising, and Fibonacci states. 

By independently measuring the diagonal elements of $\mathcal{S}$ we can thus subtract them from (\ref{eq:off-diagonal})
to get the real part of the off-diagonal elements.  Similarly, if one prepare the superposition of state as 
$\ket{\psi'}=\frac{1}{\sqrt{2}}[\ket{a}_{\beta} +i \ket{b}_{\beta}]$, one can extract the imaginary part of the 
off-diagonal elements, namely $\text{Im}\left[{_{\beta}}\boket{a}{\mathcal{S}}{b}_{\beta} \right]$. 

For generic phases where anyons are not all self-conjugate (i.e. $\mathcal{C} \neq \mathbb{I}$ ), where $\mathbb{I}$ is the identity, one 
can first measure all the matrix elements of the charge conjugation operator, ${_\beta}\boket{a}{\mathcal{C}}{b}_\beta$.   
Note the charge conjugation can also be expressed as transversal SWAPs, e.g., as $\mathcal{C} = \overline{\text{SWAP}}(1,3)(2,4)$ 
for both types of 4-layer systems in Fig.~\ref{fig:genon_protocol} (see Appendix \ref{append:loop-relation} for details).  Therefore, it can also be measured through measuring the 
SWAP operators.   Moreover, since $\mathcal{C}$ is Hermitian ($\mathcal{C}$$=$$\mathcal{C}^\dag$),  the off-diagonal elements 
can be measured by preparing the two types of superposition mentioned above.    Therefore, one can first measure 
$\boket{\psi}{\mathcal{S}}{\psi}$ as shown by Eq.~\eqref{eq:off-diagonal} and use the fact that 
\begin{align}
\non {_{\beta}}\boket{a}{(\mathcal{S}+\mathcal{S}^\dag)}{b}_{\beta}=&{_{\beta}}\boket{a}{(\mathbb{I}+\mathcal{C})\mathcal{ S}}{b}_{\beta} \\
 =& \sum_c {_\beta}\boket{a}{(\mathbb{I}+\mathcal{C})}{c}_\beta {_\beta}\boket{c}{\mathcal{ S}}{b}_{\beta}.
\end{align}
Since all the matrix elements ${_\beta}\boket{a}{\mathcal{C}}{c}_\beta $ are already measured, 
one can hence infer ${_\beta}\boket{c}{\mathcal{ S}}{b}_{\beta}$ by measurements with the 
$\frac{N(N+1)}{2}$ possible combinations of $a$ and $b$ and solving the resulting $\frac{N(N+1)}{2}$ linear 
equation, where $N$ is the number of topological sectors. 

A similar procedure can be used to measure the off-diagonal elements of other modular matrices. 
We note that since $\mathcal{R}_\alpha \mathcal{S}$ and $\mathcal{T}\mathcal{R}_\beta $ 
can be measured transversally with a 4-layer non-chiral system shown in Fig.~\ref{fig:genon_protocol}, we can 
combine the results from these measurements with the results from transversal measurements of $\mathcal{R}_\alpha$ 
and $\mathcal{R}_\beta$ in the above way, and hence extract $\mathcal{S}$ and $\mathcal{T}$ matrices.

Now we discuss possible state preparation protocols. We can start by considering a system 
with no genons (twist defects), and then adiabatically pulling out pairs of genons from the ground state.
Due to topological charge conservation, the Wilson loops surrounding each branch cut will be in the vacuum sector,
i.e.~$\ket{\mathbb{I}}_{\alpha}$ (zero anyon charge when measured along the dual $\beta$-cycle).   By applying a modular 
$\mathcal{S}$ (or $\mathcal{R}_{\alpha}\mathcal{S}$), one can also rotate the vacuum state into the dual basis, i.e., $\ket{\mathbb{I}}_{\beta}=\ket{\mathbb{I}}_{-\beta}=\mathcal{S}\ket{\mathbb{I}}_{\alpha}=\mathcal{R}_{\alpha}\mathcal{S}\ket{\mathbb{I}}_{\alpha}$. 

In order to prepare the state of a particular anyon sector $\ket{a}_{\beta}$, or superpositions of such anyon sectors, we need additional operations. 
This can be done in multiple ways. For example, one can use modular transformations themselves  as gates (see Table \ref{table:transversal_gate}) to prepare non-trivial linear combinations
of the states $\ket{a}_{\beta}$. Which particular superpositions can be applied then depends on the properties of the modular transformations; for example,
for the Fibonacci phase, arbitrary states can be prepared for genus $g \geq 2$ due to the dense covering of representations of the MCG in that case. 
For $\nu=1/k$ FQH states, the MCG generates a (generalized) Clifford group and is hence sufficient for measuring all modular matrix elements. 
In the concrete example of $\nu=1/2$ bosonic FQH state, the 2-fold degenerate ground space is equivalent to a single logical qubit. One can 
use $\mathcal{S}$ as Hadamard $\overline{H}$ and $\mathcal{T}$ as phase $\overline{P}$ (see Appendix \ref{append:fault-tolerant}) to prepare the states $\ket{0}_{\beta}$, $\ket{1}_{\beta}$, $\frac{1}{\sqrt{2}}(\ket{0}_{\beta}+\ket{1}_{\beta})$ and $\frac{1}{\sqrt{2}}(\ket{0}_{\beta}+i\ket{1}_{\beta})$ to extract all the diagonal and off-diagonal matrix elements as discussed above. 

An alternative way to prepare states is through flux insertion. We use the $\mathbb{Z}_2$ spin liquid as a concrete example, which has 
four anyon charge sectors along a particular cycle: $\ket{a}_{\beta}$, where $a = \mathbb{I}$ (vacuum), $e$ (bosonic spinon), $m$ (vison), and $em$ (fermionic spinon). 
Since the spinon carries spin-1/2, it follows that it acquires a $-1$ phase upon encircling $2\pi$ spin flux along a particular direction in spin space.
One can then show, therefore, that $\ket{m}_{\beta}=\mathcal{F}^\beta(2\pi) \ket{\mathbb{I}}_{\beta}$, where $\mathcal{F}^\beta(2\pi) $ is the operator
that adiabatically threads $2\pi$ spin flux along the $\beta$ loop. The spin flux can be taken to be, for example, along the $S_z$ spin axis. 

By inserting fractional flux, it may also be possible to create superpositions of anyons sectors; we leave further study of this for future work.

\section{Implementation of Control-SWAP}\label{append:CSWAP}

\begin{figure}
\includegraphics[width=1\columnwidth]{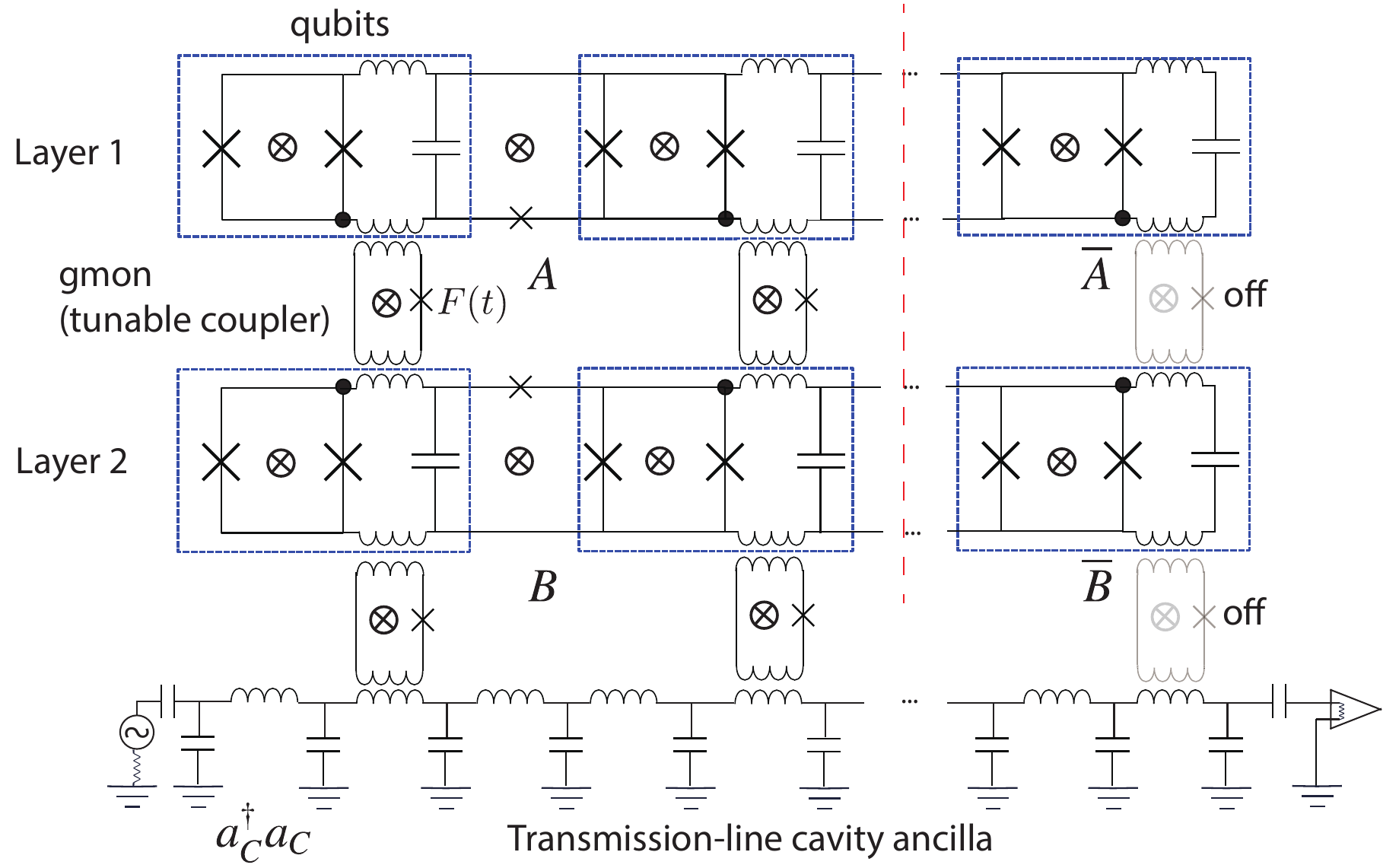}
\caption{Implementation of cavity-controlled-SWAP for the ancilla-based Ramsey interferometry using the circuit QED architecture. 
Two layers of qubit arrays are illustrated here. Each blue rectangle encircles one qubit; patches A and B ($\overline{A}$ and $\overline{B}$) 
include qubits to the left (right) of the red dashed line, in layer 1 and 2, respectively. 
The qubits in the lower layer are coupled to the cavity ancilla. All the couplings in this 
design are through the flux-tunable couplers.  This allows time-dependent control of tunneling, which enables SWAP 
and beam-splitter operations between patches $A$ and $B$ (while $\overline{A}$ and $\overline{B}$ are untouched), as 
well as the dispersive interaction between qubits in patch B and the cavity ancilla. }
\label{fig:cQEDdesign}	
\end{figure}

\nin The ancilla-based Ramsey interferometry measurement of modular matrices discussed above utilizes a Control-SWAP operation. Here we briefly review
a physical implementation of Control-SWAP in a circuit QED architecture, as shown in Fig.~\ref{fig:cQEDdesign}. An alternative implementation using a Rydberg
ancilla in a cold atom setup can be found in Ref.~\cite{Pichler:2016ec}. 

We illustrate the setup in Fig.~\ref{fig:cQEDdesign} using two layers of qubit arrays, with the lower layer coupled to a transmission-line cavity ancilla with flux-tunable 
inductive couplers, enabling local selectivity.  The qubits within and between layers are also coupled with flux-tunable inductive 
coupler (gmon)\cite{Chen:2014cw}, which enables locally selective and time-dependent tunneling to implement SWAP and beam-splitter operations.

The Control-SWAP operation can be generated by turning on the dispersive interaction between the ancilla cavity 
and the bosons (qubits) in the anti-symmetric mode after the beamsplitter operation between patch A and B
enabled by selective tunneling.  The dispersive interaction is described by the Hamiltonian \cite{Jiang:2008gs, Pichler:2016ec, Zhu:2017vi}
\be
H_\text{dis}= \chi a_C^\dag a_C \sum_j \tilde{n}_{j, B},
\ee
where $a_C^\dagger a_C = n_C$ is the number operator for the ancilla cavity mode and $\tilde{n}_{j, B}$ is the number 
operator for the anti-symmetric mode between vertically neighboring qubits from layer A and layer B. 
This effective Hamiltonian can for example be derived from the Jaynes-Cummings model \cite{Jaynes:1963fa}, which describes the interaction between cavity photons and qubits. 
In terms of the Jaynes-Cummings parameters, $\chi=g^2/\Delta$, where $g$ is the Jaynes-Cummings interaction strength 
and $\Delta=\epsilon-\omega$ is the detuning between qubit frequency ($\epsilon$) and cavity frequency ($\omega$).   
When turning on the dispersive interaction for a period $\tau = \pi/\chi$, one gets the global Control-SWAP (Fredkin) gate:
 \begin{align}
 \nonumber \text{C-SWAP} =& e^{-i H_\text{dis} \tau} = e^{-i\pi a_C^\dag a_C \sum_{j\in AB} \tilde{n}_{j,B}} = \big[ \prod_j e^{-i\pi \tilde{n}_{j, B}}  \big]^{n_C}   \\
 		       = & \big[ \overline{\text{SWAP}}  \big]^{n_C} =\mathbb{I} \otimes \ketbra{\textbf{0}_A} + \overline{\text{SWAP}} \otimes \ketbra{\textbf{1}_A}.		
 \end{align}
The SWAP operation here is controlled by the ancilla cavity photon state.  
The implementation of the dispersive interaction between the superconducting cavity ancilla and qubits can be found in Ref.~\cite{Zhu:2017vi}.

\end{appendix}


\begin{thebibliography}{55}
\bibitem{footnote1}
Note that, for simplicity, here we use the same notation $\mathcal{S}$ and $\mathcal{T}$ for both modular transformations and their representations.
\end{thebibliography}

\end{document}